\documentclass[preprint,3p,12pt]{elsarticle}
\usepackage{color}
\usepackage{amsfonts}
\usepackage{amsmath}
\usepackage{graphicx}
\usepackage{epstopdf}
\usepackage{fouriernc}
\usepackage{amssymb}
\journal{ }
\begin{document}
\begin{frontmatter}
\title{Enhanced reaction kinetics and reactive mixing scale dynamics in mixing fronts under shear flow for arbitrary Damk\"{o}hler numbers }
\author[1]{Aditya Bandopadhyay}
\ead{aditya.bandopadhyay@univ-rennes1.fr}
\author[1]{Tanguy Le Borgne}
\author[1]{Yves M\'eheust}
\address[1]{Universit{\'e} de Rennes 1, CNRS, G{\'e}osciences Rennes UMR 6118, 35041 Rennes, France}
\author[2]{Marco Dentz}
\address[2]{Spanish National Research Council (IDAEA-CSIC), E-08034 Barcelona, Spain}
\begin{abstract}
Mixing fronts, where fluids of different chemical compositions mix with each other, are known to represent hotspots of chemical reaction in hydrological systems. These fronts are typically subjected to velocity gradients, ranging from the pore scale due to no slip boundary conditions at fluid solid interfaces, to the catchment scale due to permeability variations and flow line geometries. A common trait of these processes is that the mixing interface is strained by shear. Depending on the P{\'e}clet number $Pe$, which represents the ratio of the characteristic diffusion time to the characteristic advection time, and the Damk{\"o}hler number $Da$, which represents the ratio of the characteristic diffusion time to the characteristic reaction time, the local reaction rates can be strongly impacted by the dynamics of the mixing interface. This impact has been characterized mostly either in kinetics-limited or in mixing-limited conditions, that is, for either very low or very high $Da$. Here the coupling of shear flow and chemical reactivity is investigated for arbitrary Damk{\"o}hler numbers, for a bimolecular reaction and an initial interface with separated reactants. Approximate analytical expressions for the global production rate and reactive mixing scale are derived based on a reactive lamella approach that allows for a general coupling between stretching enhanced mixing and chemical reactions. While for $Pe<Da$, reaction kinetics and stretching effects are decoupled, a scenario which we name "weak stretching", for $Pe>Da$, we uncover a "strong stretching" scenario where new scaling laws emerge from the interplay between reaction kinetics, diffusion, and stretching. The analytical results are validated against numerical simulations. These findings shed light on the effect of flow heterogeneity on the enhancement of chemical reaction and the creation of spatially localized hotspots of reactivity for a broad range of systems ranging from kinetic limited to mixing limited situations. 
\end{abstract}

\begin{keyword}
Reactive front, mixing, arbitrary Damk\"ohler number, shear flow, reaction width
\end{keyword}
\end{frontmatter}

\section{Introduction}
	Reaction fronts where two reactive fluids displace one another play an important role in a range of applications, including contaminant plume transport and reaction, soil and aquifer remediation, CO$_2$ sequestration, clogging of geothermal dipoles and the development of hotspots of reaction in mixing zones \cite{TelPhysRep2005, Tartakovsky2009, Dentz2011, Saaltink2013, FuJFM2015, HidalgoGRL2015}. Mixing of reactants by a heterogeneous flows leads to the formation of geometrically-complex fronts, at which chemical reactions occur \citep{Werth2006, Sanchez-Vila2007, DeSimoni2007, Willmann2010, CirpkaValocchi2007, chiognaGRL2012,hochkit,Rajaram_hetero, 	cirpka_hetero,dentz_JFM,fu2015rock}, Such fronts are subjected to fluid deformation which increases the surface available for diffusive mass transfer thereby enhancing effective reaction rates \cite{Ottino1979, Ou1983, Finn2011, weiss_provenzale_2008, DeWitPRL2001, Anna2014,  LeBorgne2014}.
	
	The dynamics of reactive mixing systems has been widely studied in the absence of velocity gradients. Galfi and Racz \cite{Galfi1988} and Larralde et al. \cite{Larralde1992}  studied diffusion coupled to the reversible bimolecular reaction $A + B \rightleftharpoons C$ for the case of initially well separated reactants at different bulk concentrations. The mathematical insights obtained through the long-time asymptotics are that the mass of product formed, $m_c$, grows as $m_c \sim t^{1/2}$, which is expected from the diffusive flux across the interface, while the width of the reaction front $s_r$ grows as $s_r \sim t^{1/6}$, due to the balance between the diffusive growth and the reactive consumption. For a similar reactive front, Arshadi and Rajaram \cite{Arshadi2015} found that the growth rate of the total product mass behaves as $d m_c /dt \sim t^{1/2}$ at short times while at larger times it evolves as $d m_c / dt \sim t^{-1/2}$; the transition time between the two regimes is shown to depend on the rate constant and diffusion coefficient. Similar observations of  the kinetic diffusive regimes were observed in the work by Chopard et al. \cite{Chopard1993} who focused on quantification of the influence of the reversible reaction in comparison to the forward reaction, with an emphasis on the formalism of cellular automata. Several other works relate to reaction-diffusion waves in autocatalytic systems \cite{Merkin1989}. Havlin et al. \cite{Havlin1995} and Bazant and Stone \cite{Bazant2000} have also considered the scenario of diffusion-reaction kinetics for a system with one static component (for example, a solid porous catalyst). Through their mathematical analysis, Taitelbaum and Koza \cite{Taitelbaum1998} have shown that the reaction front may move forward or backward depending upon the relative diffusivity of the two species. Benson et al. \cite{Benson_anni} have analyzed an annihilation to study the mixing density of reactants.
	When velocity gradients exist in a fluid flow, transported reactive mixtures are submitted to repeated stretching actions that lead to the formation of elongated lamellar structures. The latter are known to promote mixing and enhance reaction rates \cite{Ottino1979, Allegre1986, Rhines1983, Villermaux2000, Neufeld2009, Meunier2010}. This problem was studied by Ranz \cite{Ranz1979}, who showed that the coupling of lamella deformation with diffusion can be reduced to a 1D diffusion reaction-diffusion equation by making use of (i) the rescaled coordinate perpendicular to the direction of lamella elongation and (ii) the so-called warped time that rescales temporal increments with the lamella elongation to eliminate the stretching term. Qualitative insights about the coupling of stretching-enhanced mixing and chemical reactions were obtained in \cite{Ou1983, Clifford1999, Clifford1998} based on numerical simulations. Le Borgne et al. \cite{LeBorgne2014} investigated the impact of non-uniform flow conditions on the mixing and reaction rates under such conditions using a lamellar mixing front approach. Paster et al. \cite{PasterPRE2015} investigated the impact of shear upon reaction for uniformly initial distribution of reactants with concentration fluctuations. 

	Fluid stretching has been shown to play a fundamental role for governing mixing in porous media \cite{Villermaux2012,Battiato2009,battiato2011,boso2013,hyman2012heterogeneities}. The presence of heterogeneity in the advective flow field invariably leads to lamella formation and subsequent coalescence \cite{LeBorgne2013,LeBorgne2015,Rolle2014,bolster2016particle}. Therefore, understanding the interaction between the invading fluid and the residing fluid is imperative towards prediction of species transport in such media. For example, Mays and Neupauer \cite{Mays2012} have demonstrated a methodology to achieve enhanced mixing inducing chaotic Darcy-like flow patterns. Gramling et al. \cite{Gramling2002} performed an experimental and theoretical Darcy scale study of the reaction rate and moving front width in a porous medium for the irreversible bimolecular reaction $A+B
\rightarrow C$. They provided hints that incomplete pore-scale mixing was limiting local reaction rates, based on the inability of a Darcy scale modeling approach to properly predict the longitudinal concentration profiles and total mass of product. Later, de Anna et al. \cite{Anna2014} investigated the impact of pore-scale mixing on chemical reactions in a two-dimensional (2D) porous medium consisting of cylindrical grains allowing to measure the 2D concentration field at the pore scale. They confirmed the role of mixing in controlling local reaction rates and were able to quantitatively predict the time evolution of the measured product mass in the infinite Damk{\"o}hler limit (i.e for fast reactions) using an upscaling theory based on the concept of lamellar reaction front \cite{Jimenez-Martinez2015}. Oates \cite{Oates2007} and later Chiogna and Bellin \cite{Chiogna2013} used a concentration probability density function (PDF) based mixing model to quantify the observed reaction rates in the experimental setup of Gramling et al. \cite{Gramling2002}. 

	In this work, we focus on reactive mixing of the bimolecular reaction $A+B\rightarrow C$ under shear flow, which represents a fundamental fluid deformation and reaction process in porous media. At pore scale, shear is created by strong velocity gradients near fluid-solid interfaces \cite{Anna2014}. At Darcy scale, it is produced by spatial fluctuations in permeability \cite{LeBorgne2010,LeBorgne2015}. This broad range of scales and the diversity of chemical reactions entails the need for accounting for a wide range of reaction time scales compared to the transport time scales, that is, for a wide range of Damk{\"o}hler numbers. Therefore, we investigate analytically and numerically the global reaction kinetics of a mixing front with two initially separated diffusive reactants subjected to velocity gradients of different magnitudes, for an arbitrary finite Damk{\"o}hler number. In particular we are interested (i) in the rate at which the product is formed and (ii) in the temporal evolution of the width of the reaction front. From the analysis of asymptotic behaviors of pertinent equations we establish the existence of three distinct regimes that result from the interplay between reaction, diffusion and stretching. We pinpoint the transition times which demarcate the various regimes. We validate the analytical predictions derived for the temporal scalings of the product mass and the reaction mixing scale, i.e. the width of the reaction front, by comparison with numerical simulations. 

\section{Theoretical framework}
	
	\subsection{System description and governing equations}

\begin{figure*}[htb]
	\includegraphics[scale=0.4]{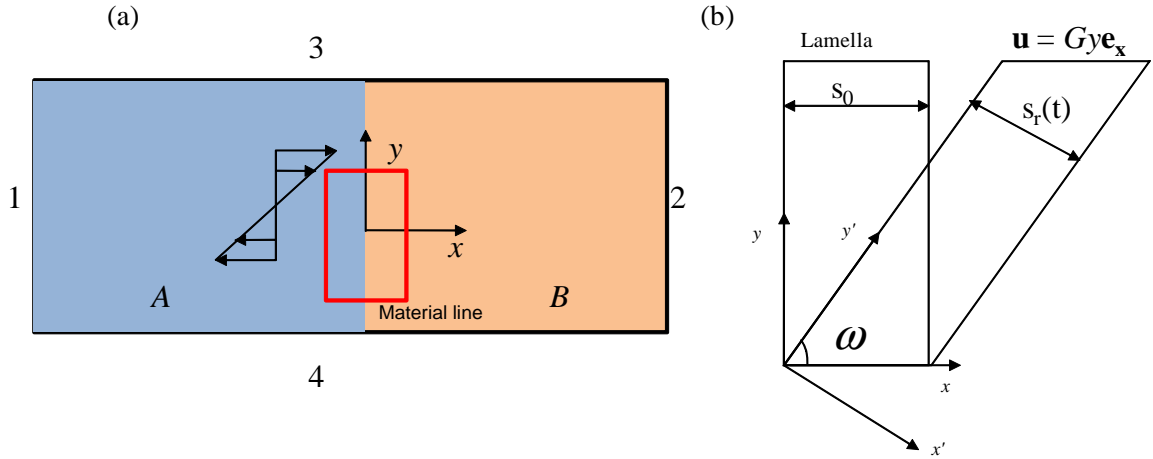}
	\begin{centering}
	\caption{General schematic and description of system employed for a 2D simulation of reaction-diffusion in the presence of a linear shear flow. (a) Boundaries 1 and 2 are considered to be periodic. Boundaries 3 and 4 are assigned to be no-flux for the 2D simulation. (b) Owing to the action of the shear flow, a material line of initial width $s_0$  deforms in the flow field so that its width reduces and becomes $w(t)$. Considering a material line positioned at the center of the mixing zone in (a), we link the transport equations defined in the 2D system presented in (a) to a 1D transport process defined in the reference frame normal to the direction of deformation of the material line. The evolution of the system is then described as a function of a 1D coordinate $x'$ which is normal to the  stretching interface.}
	\end{centering}
	\label{Fig:fig1}
	\end{figure*}

	We consider the scenario of two reactants $A$ and $B$ that are initially separated by a sharp interface. The system is subjected to a linear shear flow perpendicular to the initial interface characterized by a constant velocity gradient as shown in figure ~\ref{Fig:fig1}.a. The designated mass of reaction product grows and spreads owing to (a) diffusion, (b) advection (imposed shear flow) and (c) reaction at the interface between the two reactants. The rate of reaction is assumed to follow a second order kinetics given by $A + B\xrightarrow{k}C$ with $k$ denoting the rate constant of the chemical reaction. We consider the following two coordinate systems illustrated in figure ~\ref{Fig:fig1}.b. The coordinate system $(x,y)$  is fixed in the laboratory frame while the frame $(x',y')$ is attached to the stretching interface so that the coordinate $x'$ always corresponds to the direction perpendicular to the interface. The shear flow in the laboratory frame is denoted by ${\bf{u}} = (Gy, 0)$. Owing to the action of the shear, the front stretches about the origin as the pivot. As stretching occurs along the interfacial direction, the thickness in the direction normal to the front decreases due to the constraint of continuity (that is, the fluid mass conservation). In other words, stretching of the fluid along the interface is accompanied by a compression in the perpendicular direction of the interface. 
	
	The transport of the reactant and product concentrations, $\tilde{a}$, $\tilde{b}$ and $\tilde{c}$ respectively (the symbols with tilde represent dimensional quantities) is described by the following set of advection-reaction-diffusion equations
	
	\begin{equation}
	 \begin{aligned}
  \frac{{\partial \tilde a}}
{{\partial \tilde t}} + {\mathbf{\tilde u}}\cdot\nabla \tilde a & = {D_a}{\nabla ^2}\tilde a - k\,\tilde a\tilde b, \hfill \\
  \frac{{\partial \tilde b}}
{{\partial \tilde t}} + {\mathbf{\tilde u}}\cdot\nabla \tilde b & = {D_b}{\nabla ^2}\tilde b - k\,\tilde a\tilde b, \hfill \\
  \frac{{\partial \tilde c}}
{{\partial \tilde t}} + {\mathbf{\tilde u}}\cdot\nabla \tilde c & = {D_c}{\nabla ^2}\tilde c + k\,\tilde a\tilde b, \hfill \\ 
\end{aligned}  
	\label{eq:eqn1}
	\end{equation}
	where $D_i$, $i\in\{A,B,C\}$, represents the diffusion coefficient of the species $i$.
	
	Equation \eqref{eq:eqn1} is subjected to the following initial conditions: 
	\begin{equation}
\begin{gathered}
  \tilde a\left( {{\mathbf{\tilde x}},0} \right) = \left \arrowvert {\begin{array}{*{20}{c}}
   {{{\tilde n}_0}\,{\text{~ if ~}}\,\tilde x < 0,\forall \tilde y}  \\
   {0\,\,\,{\text{otherwise}}}  \\

 \end{array} } \right.{\text{,}} \hfill \\
  \tilde b\left( {{\mathbf{\tilde x}},0} \right) = \left \arrowvert  {\begin{array}{*{20}{c}}
   {0\,\,\,{\text{~ if ~}}\tilde x \geq 0,\forall \tilde y}  \\
   {{{\tilde n}_0}\,\,\,{\text{otherwise}}}  \\

 \end{array} } \right.{\text{,}} \hfill \\
  \tilde c\left( {{\mathbf{\tilde x}},0} \right) = 0\,\,\forall {\mathbf{\tilde x}}{\text{ }}{\text{.}} \hfill \\ 
\end{gathered} 
	\label{eq:bcs}
	\end{equation}
The velocity field is given by ${\bf{u}} = Gy {\bf{e}}_x$ with $G$ representing the shear rate. The boundary conditions for the species $A$ and $B$ are written down keeping in mind that the widths are infinitely long. There is, thus, no specific length scale attached to the problem; this is akin to most semi-infinite problems for diffusion, such as observed in heat transfer and mass transfer in a semi-infinite domain \cite{Crank1975, incropera1985}). In order to address the problem computationally for the 2D simulations, we make the boundary conditions at the left and right boundaries periodic, where the domain size is selected in such a way that over the time of observation, the system does not experience the effects of the boundaries.

	
	In this work, we assume for simplicity that the diffusion coefficients of all the participating species are equal and denoted $D$. We nondimensionalize equation \eqref{eq:eqn1} by the following quantities: ${\bf{x}} = {\bf{\tilde{x}}/\tilde{\delta_0}}$ (in the system under consideration we choose the characteristic length to be of  $\delta_0$), $t = \tilde{t}D/\tilde{\delta_0^2}$ (the characteristic time scale of the system based on the characteristic length, $\tilde{\delta}_0^2/D$), ${\bf{u}} = {{\bf{\tilde{u}}}/G\tilde{\delta_0}}$ (the characteristic velocity of the system is proportional to the rate of shear times the characteristic length), $\phi = \tilde{\phi}/\tilde{n}_0$ where $\phi = a,b,c$. Proceeding based on this nondimensionalization, and defining the P{\'e}clet and Damk{\"o}hler numbers as, 
	\begin{equation}
	\begin{aligned}
Pe = \frac{\tau_D}{\tau_a}=\frac{G\tilde{\delta}_0^2}{D}, \\
Da = \frac{\tau_D}{\tau_r}=\frac{k\tilde{n_0}\tilde{\delta_0}^2}{D}, \\ 
\end{aligned} 
	\label{eq:Pe:Da}
	\end{equation}	
which represent respectively the ratio of the typical diffusion time scale $\tau_D=\tilde{\delta}_0^2/D$ to the typical advection time scale $ \tau_a= G^{-1}$ and that of the typical diffusion time scale to the typical reaction time scale respectively $\tau_r=({k\tilde{n_0}})^{-1}$, equations \eqref{eq:eqn1} and \eqref{eq:bcs} may be then recast as: 
	\begin{equation}
	\begin{aligned}
\frac{{\partial a}}{{\partial t}} + Pe\: {\mathbf{u}}\cdot\nabla a & = {\boldsymbol \nabla ^2}a - Da\,a\, b, \\
  \frac{{\partial b}}
{{\partial t}} + Pe\: {\mathbf{u}}\cdot\nabla b & = {\boldsymbol \nabla ^2}b - Da\,a\, b, \\
  \frac{{\partial c}}
{{\partial t}} + Pe\: {\mathbf{u}}\cdot\nabla c & = {\boldsymbol \nabla ^2}c + Da\,a\, b \hfill, \\ 
\end{aligned} 
	\label{eq:nd1}
	\end{equation}
	subjected to 
	\begin{equation}
	\begin{aligned}
 a\left( {{\mathbf{x}},0} \right) & = \left\{ {\begin{array}{*{20}{c}}
   {1\,\,\,{\text{if  }}x < 0,\forall y}  \\
   {0\,\,\,{\text{otherwise}}}  \\

 \end{array} } \right.{\text{  }}{\text{,   }} \hfill \\
  b\left( {{\mathbf{x}},0} \right) & = \left\{ {\begin{array}{*{20}{c}}
   {0\,\,\,{\text{if  }}x \geq 0,\forall y}  \\
   {1\,\,\,{\text{otherwise}}}  \\

 \end{array} } \right.{\text{  }}{\text{,   }} \hfill \\
  c\left( {{\mathbf{x}},0} \right) & = 0\,\,\,\,\forall {\mathbf{x}}{\text{ }}{\text{.}} \hfill \\ 
\end{aligned} 
	\label{eq:ndbcs}
	\end{equation}
As we shall see later, depending on the relative strengths of $Pe$ and $Da$ the system evolution exhibits markedly different dynamics.

Note that the scenario where there is no definite length scale associated with the particular problem under consideration can be tackled in the following manner. A system lacking a characteristic lengthscale is akin to analysis pertaining to semi-infinite domains where the pertinent lengthscale must be obtained by a combination of the physically relevant parameters in the system \cite{kundu2008fluid}. In the context of the present problem, we may define two length scales based on the combination of the diffusion coefficient, $D$, the reaction kinetics constant, $k$, and the average shear rate, $G$. If the system is thought to evolve based on the balance of diffusion and reaction kinetics the length scale $\tilde{\delta}_0$ may be recast as $\sqrt{D/(kn_0)}$, which represents the characteristic distance of diffusion $\sqrt{D \tau_r}$ over the the characteristic reaction time $\tau_r=1/(kn_0)$. In the other case, if the system evolves based on the balance of diffusion and stretching, the lengthscale $\tilde{\delta}_0$ may be recast as $\sqrt{D/G}$, which quantifies the characteristic diffusion scale over the characteristic stretching time $\tau_s=1/G$. In the present derivation, we do not restrict ourselves to the choice of $\tilde{\delta}_0$ thereby allowing us to generalize the analysis presented here. While presenting the results we shall also dwell on the two aforementioned choices of typical lengths, and quantify the observations from the respective physical viewpoints.

	\subsection{Reaction-diffusion system in a Lagrangian frame}
	\label{sec:system_descrip}
To derive analytical expressions quantifying the coupling between reaction, diffusion and shear , we attempt to simplify the equations and make them more tractable by reducing the dimensionality of the problem \cite{Ranz1979, Alvarez1998,Cerbelli2002, Meunier2010}. In order to provide a better insight into the fundamental process of stretching augmented reaction-mixing, we convert the 2D transport problem defined in the laboratory frame to a one-dimensional (1D) transport problem defined in the local Lagrangian frame attached to a particular material line. A material line is a purely kinematic quantity, independent of the species residing in it \cite{Ranz1979}, but we choose the material line that coincides with the middle line of the mixing zone (see Figure~\ref{Fig:fig1}), which is initially oriented along the direction $y$. Considering a volume of initial constant thickness ${\tilde \delta }_0$ around the material line (also referred to as a lamella, see figure \ref{Fig:fig1}), the elongation of the line by shear deformation $\tilde \rho(t)$ leads to a simultaneous compression of the thickness $\tilde \delta$ (owing to incompressibility), such as \cite{Ranz1979,Meunier2010},   
%
%
%
		\begin{equation}
		\tilde \rho = {{\sqrt {1 + {G^2}{\tilde t^2}} }} \text{~,}
		\label{eq:k1:rho}
		\end{equation}
and
		\begin{equation}
		\frac{{\tilde \delta }}
{{{{\tilde \delta }_0}}} = \frac{1}
{{\sqrt {1 + {G^2}{\tilde t^2}} }} \text{~.}
		\label{eq:k1}
		\end{equation}
The temporal relative rate of compression due to stretching may be found out by differentiating equation ~\eqref{eq:k1} as 
\begin{equation}	 
\frac{1}
{{\tilde \delta }}\frac{{d\tilde \delta }}
{{d\tilde t}} = \frac{{ - {G^2}\tilde t}}
{{1 + {G^2}{{\tilde t}^2}}} = -\Omega \text{~.}
\label{eq:kd1}
\end{equation}
Upon shifting the point of view to the Lagrangian frame $(x',y')$, the local velocities are $v'_x = -\Omega x', v'_y = \Omega y'$, which clearly indicates that the lamella is being compressed in the $x'$ direction while simultaneously being elongated in the $y'$ direction. For this particular system, let us write down the governing transport equation for the species $a$ (the governing equations for the other species will be analogous to this): 

\begin{equation}
\frac{{\partial \tilde a}}
{{\partial \tilde t}} + {\mathbf{\tilde u'}}\cdot\nabla \tilde a = D{\nabla ^2}\tilde a - k \tilde{a}\tilde{b} \text{~,}
\label{eq:lag1}
\end{equation}
where $\tilde{u'}$ denotes the velocity in the Lagrangian frame. The reaction rate is assumed to be given by a simple second order kinetics $k \tilde{a} \tilde{b}$. As can be seen from equation \ref{eq:k1}, with progress in time, the linear stretching makes the thickness of the width of the material line decrease rapidly and it so happens that the aspect ratio of the structure becomes small. Moreover since the concentration along the $y'$ direction does not vary as well as the fact that concentration gradient along the material line is zero, the gradients in the $y'$ direction may be neglected as compared to those in the $x'$ direction\cite{Ranz1979,Meunier2010}. 

Therefore, the simplified equation with the aforementioned simplification may be written as
\begin{equation}
\frac{{\partial \tilde a}}
{{\partial \tilde t}} - \Omega \tilde x'\frac{{\partial \tilde a}}
{{\partial \tilde x'}} = D\frac{{{\partial ^2}\tilde a}}
{{\partial \tilde x{'^2}}} + \tilde R \tilde{~,}
\label{eq:redlag1}
\end{equation}
We have thus successfully simplified the 2D set of equations \eqref{eq:lag1} by neglecting the axial gradients to obtain equation \eqref{eq:redlag1}, which is an advection-diffusion-reaction system in 1D. We can make further progress by transforming the above equation into a diffusion-reaction system by making use of the following rescaled variables \cite{Ranz1979}: 
\begin{equation}
z = \frac{\tilde x'}{\tilde{\delta}}\text{~ and ~}\theta  = \int_0^{\tilde{t}}d\tau\: \frac{D}{\tilde{\delta}(\tau)^2} \text{~.}
\label{eq:warpedvar}
\end{equation}
where $\theta$ is referred to as the warped time which represents the integral diffusion time over the lamella thickness. We obtain-
\begin{equation}
\begin{gathered}
  \frac{{\partial \bar a}}
{{\partial \theta }} = \frac{{{\partial ^2}\bar a}}
{{\partial {z^2}}} - k\bar a\bar b\frac{{{{\bar \delta }^2}}}
{D}, \hfill \\
  \frac{{\partial \bar b}}
{{\partial \theta }} = \frac{{{\partial ^2}\bar b}}
{{\partial {z^2}}} - k\bar a\bar b\frac{{{{\bar \delta }^2}}}
{D} \hfill, \\
  \frac{{\partial \bar c}}
{{\partial \theta }} = \frac{{{\partial ^2}\bar c}}
{{\partial {z^2}}} + k\bar a\bar b\frac{{{{\bar \delta }^2}}}
{D} \hfill. \\ 
\end{gathered} 
\label{eq:warpeqn}
\end{equation}
Furthermore, a nondimensionalization may be employed for the lamella thickness as $\delta = \tilde{\delta}/\tilde{\delta}_0$, and by taking advantage of equation~\eqref{eq:kd1} we may simplify the definition of the warped time to obtain the form
\begin{equation}
\theta = \int_0^{\tilde{t}}\frac{d\tau}{\tilde{\delta}\left(\tau\right)^2/D} = \frac{D}{\tilde{\delta}(\tau)} \int_0^{\tilde{t}}d\tau \left(1+G^2\tilde{t}^2\right) = \frac{D}
{{\tilde \delta _0^2}}\left( {\tilde t + \frac{{{G^2}{{\tilde t}^3}}}
{3}} \right)  \text{~ .}
\label{eq:warped}
\end{equation}
 Using this, and the aforementioned nondimensional scheme, we may write the governing equations in the Lagrangian frame for the three species in the frame attached with the deforming material line as  
 \begin{equation}
\begin{gathered}
  \frac{{\partial a}}
{{\partial \theta }} = \frac{{{\partial ^2}a}}
{{\partial {z^2}}} - Da\,ab{\delta ^2}, \hfill \\
  \frac{{\partial b}}
{{\partial \theta }} = \frac{{{\partial ^2}b}}
{{\partial {z^2}}} - Da\,ab{\delta ^2}, \hfill \\
  \frac{{\partial c}}
{{\partial \theta }} = \frac{{{\partial ^2}c}}
{{\partial {z^2}}} + Da\,ab{\delta ^2}, \hfill \\ 
\end{gathered}  
 \label{eq:warpedeq}
 \end{equation}
 where the P{\'e}clet number and Damk{\"o}hler numbers have already been defined in section \ref{sec:system_descrip} and
 \begin{equation}
 \delta  = \frac{1}{\sqrt {1 + P{e^2}{t^2}}} 
 \label{eq:def_delta}
 \end{equation}
from equation \eqref{eq:k1}).
 
 The solution for the set of equations ~\eqref{eq:warpeqn} yields the temporal and spatial evolution for the concentration fields of the different species. 
The initial conditions specified for solving equation \eqref{eq:warpeqn} are that as $z\rightarrow-\infty$, $a=1,b=0$ and that as $z\rightarrow\infty$, $b=1,a=0$. The aim of the present work is to depict how the chemical kinetics constants in conjunction with the lamella stretching impact the mass of the product formed in such a scenario. 
 
 We may write the mass of the product as $\tilde{m}_c = \int_{-\infty}^{\infty}dx'\tilde{c}(x')l(t)$, where $l(t)$ represents the length of the interface. In a nondimensional sense, it may be recast as 
 \begin{equation}
 m_c = \frac{\tilde{m}_c}{n_0 l_0 \tilde{\delta_0}} =  \int_{-\infty}^{\infty}dz\, c(z)l\delta = \int_{-\infty}^{\infty}dz\, c(z) \text{~,}
 \label{eq:mass}
 \end{equation}
 where we have made use of the fact that $l \delta = 1$ (owing to the incompressibility condition which imposes that the amount of longitudinal stretching be the inverse of the amount of transverse compression in order to conserve the material volume).
 
Another quantity of interest is the reactive mixing scale, which quantifies the spatial localization of chemical reactivity in a mixing front \cite{Larralde1992}. A small reactive mixing scale denotes the existence of localized hotspots of chemical reaction, while a large reactive mixing scale implies that the reaction zone is diffuse. The reactive mixing scale is thus an important characteristic of reactive fronts, which, as discussed in the following, depends non-trivially on $Pe$, $Da$ and on the observation time. The reactive mixing scale can be estimated from the half-width of the reaction front from the second moment of the reaction rate across the front as 
\begin{equation}
\tilde{s}_r = \left( \frac{\int_{-\infty}^{\infty}dx' \, x'^2 Da ab}{\int_{-\infty}^{\infty}dx' \, Da {a b}}\right)^{1/2}\text{~,}
\label{eq:rxnwidth}
\end{equation} 
In a nondimensional sense, it can be written in terms of the integral in the transformed coordinate, $z$, as 
\begin{equation}
s_r = \left( \frac{\int_{-\infty}^{\infty}dz\, z^2 Da {a b}}{\int_{-\infty}^{\infty}dz \, Da {a b}}\right)^{1/2} \delta \text{~.}
\label{eq:ndrxn}
\end{equation}
Note that the reactive mixing scale is different from the conservative mixing scale \cite{LeBorgnePRE2011, LeBorgne2015} since it is affected by chemical reactions that tend to reduce it.

\section{Analytical predictions for the temporal evolution of the mass of product}
Having discussed the general 1D framework used to investigate the transport of reactive species, we focus on the various limits of the temporal evolution of the mass of the product. 
We distinguish here three characteristic temporal regimes. Let us remind here that the P{\'e}clet number denotes the ratio of the typical diffusion time scale (over the distance $\delta_0$) to the typical advection time scale (due to the fluid shear), while 
 the Damk{\"o}hler number denotes the ratio of the typical reaction time scale to the typical diffusion time scale. This implies that the dimensionless time $t=\tilde{t}/\tau_D$ allows us to distinguish the dynamics of the systems in terms of the relative values of $Pe^{-1}=\tau_a/\tau_D$ and $Da^{-1}=\tau_r/\tau_D$.

The first scenario, termed as \textit{weak stretching} scenario, is defined by a system for which $Da^{-1}< Pe^{-1}$. In such a scenario, 3 different time regimes can be distinguished. The first regime occurs for $t\ll Da^{-1}$. Here the reaction dynamics is determined by the interaction of diffusion and reaction kinetics since the time is small compared to the characteristic reaction time. In the second regime, occurring for $Da^{-1}<t<Pe^{-1}$, the reaction is fully diffusion limited since the time is large compared to the reaction time (i.e. reactions are fast) but small compared to the characteristic advection time (i.e. stretching does not play a role yet). In the asymptotic long time regime, $t\gg Pe^{-1}$, the shear action of the flow field is activated and the reaction kinetics is controlled by stretching-enhanced diffusion. 

The second scenario, termed as the \textit{strong stretching} scenario is defined as $Pe^{-1}< Da^{-1}$. The first regime, characterized by $t\ll Pe^{-1}$ is similar to the first regime in the weak stretching case. The system is dominated by the interaction of diffusion and reaction kinetics. The second time regime, for which $Pe^{-1} < Da^{-1}$, sees a reaction behavior that results from the interaction of the reaction kinetics and stretching enhanced diffusion as time is large enough for stretching to be activated but small compared to the characteristic reaction time. This regime is particularly interesting as stretching and reaction kinetics are fully coupled. In the long time regime, $t\gg Da^{-1}$, the reaction behavior is fully limited by stretching-enhanced diffusion as reaction kinetics is no longer limiting. 

As detailed in the introduction, reaction front kinetics have been investigated mostly in the limits $t\ll Pe^{-1}$ (no shear) or $t\gg Da^{-1}$ (fast reactions under shear), which represents only a subset of the range of possible regimes described above. In the following we derive the temporal evolution of the product mass for all the aforementioned regimes.

	\subsection{Weak stretching scenario: $Da^{-1}< Pe^{-1}$}
	\label{sec:weakscenario}

For $t\ll Pe^{-1}$ shear does not affect the reactive transport behavior as both the elongation $\rho$ and the width $\delta$ are approximately constant (see equations \eqref{eq:k1:rho} and \eqref{eq:k1}). Thus, the evolution of the product $c$ can be represented by the diffusion reaction equation 
\begin{equation}
\frac{\partial c}{\partial t} = \frac{\partial ^2 c}{\partial x^2} + Da a b.
\label{no_shear}
\end{equation}
 Hence, this regime is akin to the case where there is no imposed flow field \cite{Larralde1992}. Conversely for $t\gg Pe^{-1}$, shear plays an important role in increasing the area available for diffusive mass transfer, and enhancing chemical gradients by compression. In the following, we discuss the three regimes separated by the characteristic times $Da^{-1}$ and $Pe^{-1}$. 
	
	\subsubsection{Reaction-diffusion regime (negligible shear): $t\ll Da^{-1}$}
	\label{sec:p1}
	In this regime, the initial profiles of $a$ and $b$ both evolve due to diffusion while the impact of reaction is still weak. In such a case, the concentration profiles of the two reactants can be approximated by the diffusive profiles \cite{Crank1975}
	\begin{eqnarray}
	\label{diffusive_profile_a}	
	a = \frac{1}{2}(1+\text{erf}(x/2\sqrt{t})), \\
	b = \frac{1}{2}(1-\text{erf}(x/2\sqrt{t})),
	\label{diffusive_profile_b}
	\end{eqnarray}
Thus, in this regime where we can neglect the effect of shear, we may write the evolution of $c$, 
inserting equation \eqref{diffusive_profile_a} and \eqref{diffusive_profile_b} in equation \eqref{no_shear}, as 
\begin{equation}
\frac{\partial c}{\partial t} = \frac{\partial ^2c}{\partial x^2} + Da\frac{1}{4} \left[ 1-\text{erf} \left ( \frac{x}{2\sqrt{t}} \right )  ^2 \right ]
\end{equation}
which, in the vicinity of the interface, i.e. the origin (where most of the product is formed) may be further simplified as 
\begin{equation}
\frac{\partial c}{\partial t} = \frac{\partial ^2c}{\partial x^2} + Da\frac{1}{4}.
\label{eq:t_0_Da}
\end{equation}
We may integrate this over space to obtain the temporal evolution of the mass of product formed as
\begin{equation}
\frac{\partial m_c}{\partial t} = \left[ \frac{\partial c}{\partial x} \right]_{-\infty}^{\infty}+ \frac{Da}{4}s_r
\label{eq:mass_Da_0}
\end{equation}
wherein the second term on the right hand side has a contribution from $Da/4$ which is concentrated near the interface (at the origin), while $s_r$ represents the reactive mixing scale. 
The first term on the right hand size is zero on account of the fact that the concentration profile decays away from the front towards either reactant so that $\partial c/ \partial x (x\rightarrow\pm\infty)= 0$. Now, at such times where $t \ll Pe^{-1}$ and $t \ll Da^{-1}$, the growth of the reactive mixing scale is only {\it{diffusion controlled}}, which leads to the growth of the reactive mixing scale as $s_r \sim \sqrt{t}$.  Thus, we arrive at the evolution of $m_c$ by integrating equation \eqref{eq:mass_Da_0} in time as 
\begin{equation}
m_c \approx \frac{Da}{4} t^{3/2}.
\label{eq:mc_Da_0}
\end{equation}

This result is equivalent to that obtained by Arshadi and Rajaram \citep{Arshadi2015} in the early time regime of a reaction front with no shear. 

\subsubsection{Diffusion limited regime (negligible shear at intermediate times): $Da^{-1}<t<Pe^{-1}$} 
	\label{sec:p3}
In order to determine the behavior at longer times (where the influence of the shear is still not present; this regime is similar to the case at long times in the absence of any imposed flow), we follow the lead of Larralde et al. \cite{Larralde1992}. For completeness, we summarize here the main steps of the derivation and the key results. We begin by representing the concentration of the product as a perturbation on the diffusive case. Formally, we first observe that the concentrations of the two reactants may be written as $a = F + g$ and $b = g$ where $F = a-b$ satisfies the conservative equation (by noting that subtracting the two concentration fields gets rid of the nonlinearity of the reaction term)

\begin{equation}
\frac{{\partial F}}
{{\partial t}} = \frac{{{\partial ^2}F}}
{{\partial {x^2}}}   \text{~ .}
\label{eq:eq_for_F}
\end{equation} 	
where $g$ represents the concentration perturbation.
Consequently, the governing equation for $g$ may be represented as
\begin{equation}
\frac{{\partial g}}
{{\partial t}} = \frac{{{\partial ^2}g}}
{{\partial {x^2}}} - Da\,g\left( {\text{erf}\left( {\frac{x}
{{\sqrt {4t} }}} \right) + g} \right)  
\label{eq:eq_n'}
\end{equation}
which may be simplified by neglecting the nonlinear contribution from $g^2$ (since $g$ is considered to be a perturbation to $F$) and linearizing $\text{erf}( {{x}/{{\sqrt {4t} }}} )$ for times so large and the point of interest such that $x/\sqrt{4t} \ll 1$, we obtain 
	\begin{equation}
\frac{{\partial g}}
{{\partial t}} \approx \frac{{{\partial ^2}g}}
{{\partial {x^2}}} - Da\,g\frac{x}
{{\sqrt {\pi t} }} \text{~ .}
	\label{eq:Ai_1}
	\end{equation}
	At times sufficiently long for the temporal derivative to be negligible, the time behaves as a parameter to the Airy differential equation given by: $\frac{{{\partial ^2}g}}
{{\partial {x^2}}} - Da\,g\frac{x}
{{\sqrt {\pi t} }} = 0$. The solution to this may be written by resorting to an Airy function, as: 
	\begin{equation}
g\sim f\left( t \right)Ai\left( {\lambda \frac{x}
{{{t^{1/6}}}}} \right),\;\lambda=\left( \frac{Da}{\sqrt{\pi}} \right)^{1/3} \text{~ .}
	\label{eq:Ai_2}
	\end{equation}
	We may assume a power-law representation for the function $f(t)$ so that the form for the concentration perturbation is represented as $g = \psi t^{\alpha} Ai (\lambda x/t^{1/6})$.  By equating the form of the nonlinear term in \eqref{eq:eq_n'} and the spatial derivative term on the right hand side of equation \eqref{eq:Ai_2}, we see that $\alpha = -1/3$ and $\psi = 1/\lambda$. Thus, the perturbation to the concentration is found out as 
	\begin{equation}
	g \sim \frac{t^{-1/3}}{\lambda}Ai\left(\frac{\lambda x}{t^{1/6}} \right) \text{~ .}
	\label{eq:conc}
	\end{equation}
	The above may be further simplified for the case of large enough $x$ such that $\lambda x \gg  t^{1/6}$ to yield
	\begin{equation}
g\sim\frac{{{t^{ - 1/3}}}}
{\lambda }{\left( {\frac{{\lambda x}}
{{{t^{1/6}}}}} \right)^{ - 1/4}}\exp \left( { - \frac{2}
{3}{{\left( {\frac{{\lambda x}}
{{{t^{1/6}}}}} \right)}^{3/2}}} \right) \text{~ ,}
	\label{eq:n'}
	\end{equation}
	while the reaction term may be recast based on Eq.\eqref{eq:Ai_1}) to yield
	\begin{equation}
R\sim Da\frac{x}
{{\sqrt t }}\frac{{{t^{ - 1/3}}}}
{\lambda }{\left( {\frac{{\lambda x}}
{{{t^{1/6}}}}} \right)^{ - 1/4}}\exp \left( { - \frac{2}
{3}{{\left( {\frac{{\lambda x}}
{{{t^{1/6}}}}} \right)}^{3/2}}} \right) \text{~ .}
	\label{eq:R}
	\end{equation}
	The time integral of the reaction term may then be obtained as $I \sim \int_{0}^{\infty}R dt$, which, according to equation~\eqref{eq:R} can be written as
	\begin{equation}
	I \sim \int_{0}^{\infty} dt Da\frac{x}
{{\sqrt t }}\frac{{{t^{ - 1/3}}}}
{\lambda }{\left( {\frac{{\lambda x}}
{{{t^{1/6}}}}} \right)^{ - 1/4}}\exp \left( { - \frac{2}
{3}{{\left( {\frac{{\lambda x}}
{{{t^{1/6}}}}} \right)}^{3/2}}} \right) \text{~ .}
	\label{eq:I}
	\end{equation}
	Upon simplification, equation \eqref{eq:I} can be written, in leading order, as 
	\begin{equation}
I\sim\lambda {t^{1/3}}{\left( {\frac{{\lambda x}}
{{{t^{1/6}}}}} \right)^{ - 3/4}}\exp \left( { - \frac{2}
{3}{{\left( {\frac{{\lambda x}}
{{{t^{1/6}}}}} \right)}^{3/2}}} \right)
	\end{equation}
Towards determining the mass of product, we appeal to the above equation and equation \eqref{eq:mass} to obtain
\begin{equation*}
m_c = 2\int_{0}^{\infty}dx\lambda {t^{1/3}}{\left( {\frac{{\lambda x}}
{{{t^{1/6}}}}} \right)^{ - 3/4}}\exp \left( { - \frac{2}
{3}{{\left( {\frac{{\lambda x}}
{{{t^{1/6}}}}} \right)}^{3/2}}} \right)
\end{equation*}	
where the factor of 2 appears when the spatial integration is done from the lower bound $0$ instead of $-\infty$ while keeping in mind that the integral is symmetric for the concentration perturbation. The integral can be evaluated by making the change of variable $\beta = \lambda x/t^{1/6}$ to obtain
\begin{equation}
m_c \approx t^{1/2}2\int_{0}^{\infty} d\beta \exp\left( \frac{2}{3}\beta^{3/2}\right)\beta^{-3/4} \approx 8t^{1/2} \text{~ .}
\label{eq:mc_3}
\end{equation}
	The general observation is that the long time temporal evolution of the mass of the product formed becomes independent of the Damk{\"o}hler number, which is expected for this regime that is purely diffusion-limited. We note here that the integral appearing in equation \eqref{eq:mc_3} comes out to be $2\int_0^{\infty}dx \exp(2x^{3/2}/3)x^{-3/4} = 2/3\times (2^{5/6}3^{1/6}\pi/)\Gamma(5/6)$  whose approximate value is $7.94$, and which we further approximate as $8$ in the equation above. This result is in line with the derivation of Larralde \cite{Larralde1992} and of Arshadi and Rajaram \cite{Arshadi2015} for a reactive front with no shear.

\subsubsection{Shear-enhanced reactive mixing regime: $t\gg Pe^{-1}$}
	\label{sec:p4}
We now turn our attention to the case where the imposed velocity gradient has significant bearing on the dynamics of the reaction-diffusion system. Towards analyzing this, we begin with equation \eqref{eq:warpeqn} and move ahead with the logic similar to that employed in section \ref{sec:p3}. In $\left\lbrace \theta, z \right\rbrace$ coordinates, which account for the effect of shear-induced elongation and compression of the interface, we may write the evolution of the perturbation to the concentrations of $a$ and $b$, as in equation \eqref{eq:eq_n'}, as 
\begin{equation}
\frac{{\partial g}}
{{\partial \theta }} = \frac{{{\partial ^2}g}}
{{\partial {z^2}}} - \frac{{Da}}
{{1 + P{e^2}{t^2}}}g\left( {\text{erf}\left( {\frac{z}
{{\sqrt {4\theta } }}} \right) + g} \right) \text{~,}
\end{equation}
which can be further simplified by neglecting the nonlinear contribution of the $g^2$ term to yield
\begin{equation}
\frac{{\partial g}}
{{\partial \theta }} = \frac{{{\partial ^2}g}}
{{\partial {z^2}}} - \frac{{Da}}
{1+{P{e^2}{t^2}}}g\frac{z}
{{\sqrt {\pi \theta } }}  \text{~,}
\label{eq:eq_4_1} 
\end{equation}
where we have made use of the fact that since we are looking at sufficiently long times, we may observe the region where $t>Pe^{-1}$, allowing us to write $1/{1+Pe^2t^2} \approx 1/Pe^2t^2$. For convenience, we consider the system in the $x'$ and $t$ coordinates which are easily transformed by recalling that $z = x'/\delta \sim x' Pe\, t$ and $\theta \sim Pe^2 t^3 / 3$, obtaining
\begin{equation}
\frac{{\partial g}}
{{\partial t}} = \frac{{{\partial ^2}g}}
{{\partial x{'^2}}} - Da\sqrt {\frac{3}
{\pi }} g\frac{{x'}}
{t} \text{~ .}
\label{eq:simplified_41}
\end{equation}
The solution to equation \eqref{eq:simplified_41} may be obtained (in analogous manner to that in section \ref{sec:p3}) as 
\begin{equation}
g \approx \frac{t^{-1/3}}{\lambda'}Ai\left( \frac{\lambda' x'}{t^{1/6}}\right) \text{~,}
\label{eq:sol41}
\end{equation}
where $\lambda' = (Da \sqrt{3/\pi})^{1/3}$.
Proceeding, we may write the rate of reaction term as 
\begin{equation}
R = Da \frac{x'}{t^{1/2}}g \approx \lambda'^3 \frac{x'}{t^{1/2}} \times \frac{t^{-1/3}}{\lambda'}Ai\left( \frac{\lambda x'}{t^{1/6}}\right) \text{~ ,}
\end{equation}
which we may simplify to obtain 
\begin{equation}
R \approx \lambda \left( \frac{\lambda x'}{t^{1/6}} \right) t^{-2/3} Ai\left( \frac{\lambda x'}{t^{1/6}} \right) \text{~ .}
\label{eq:R_Airy_function}
\end{equation}
Proceeding further, we may integrate the above expression in space and time to obtain 
\begin{equation}
m_c \approx \lambda t^{1/3} 2\int_{0}^{\infty} dz {\left( {\frac{{\lambda x}}
{{{t^{1/6}}}}} \right)^{ - 3/4}}\exp \left( { - \frac{2}
{3}{{\left( {\frac{{\lambda x}} 
{{{t^{1/6}}}}} \right)}^{3/2}}} \right) \text{~ ,}
\label{eq:m4}
\end{equation}
the evaluation of which necessitates converting the groups of the variables into the form of $\gamma = (\lambda z)(Pe\, t^{7/6})$. Performing this, we obtain from equation \eqref{eq:m4} that 
\begin{equation}
m_c \approx 8Pe\, t^{3/2} \text{~ .}
\end{equation}
This expression shows that, as expected, the impact of the shear action of the flow leads to a stronger increase in the product mass than that due to pure diffusion as given by equation \eqref{eq:mc_3}. This result is validated against numerical simulations in section \ref{sec:results}.

For the weak stretching scenario, discussed in this section, the effects of kinetics limitation, due to finite reaction times, are confined to the domain $t \ll Da^{-1}$, while the effects of mixing enhancement by shear are confined to the domain  $t \gg Pe^{-1}$. Hence, as $Da^{-1} < Pe^{-1}$, these two domains are disjoint and these two effects are decoupled. As discussed in the next section, this is not the case for the strong stretching scenario, which leads to new scaling laws.

	\subsection{Strong stretching scenario: $Pe^{-1}< Da^{-1}$}
	\label{sec:p2}
In the strong stretching scenario, the shear time scale $Pe^{-1}$ is smaller than the characteristic reaction scale $Da^{-1}$. In this scenario, the diffusion-limited reaction regime $Da^{-1}<t<Pe^{-1}$ does not exist. Instead, we find a new regime characterized by the interaction of stretching-enhanced mixing and reaction.  It must first be noted that for $t<Pe^{-1}$ the effect of stretching, as manifested through the  presence of the term $\delta^2$ in the reaction term (where we recall that $\delta = 1/\sqrt{1+Pe^2t^2}$), is weak, and thus the approximate equations assuming $1+Pe^2t^2 \approx 1$ result in the analysis seen in section \ref{sec:p1}. Essentially this implies that in the initial time, the system behaves as if there is no imposed shear. On the other hand, in the long time regime ($t\gg Da^{-1}$), the reaction behavior is the same as the one obtained in section \ref{sec:p4}, i.e. that of shear enhanced reactive mixing with no kinetics limitations (i.e. fast reactions compared to the observation time). 
	In the following, we therefore only study the reaction behavior in the intermediate regime ($Pe^{-1} < t < Da^{-1}$) in which the effect of flow stretching and kinetics limitations are both non-negligible. We thus begin the analysis through equation \eqref{eq:warpedeq} in terms of $\left\lbrace \theta, z \right\rbrace$ coordinates: 
	\begin{equation}
	\frac{\partial c}{\partial \theta} = \frac{\partial ^2c}{\partial z^2} + Da a b \delta^2 \text{~ ,}
	\label{eq:warped_c}
	\end{equation}
As in section \ref{sec:p1}, approximating the profiles of $a$ and $b$ by diffusive profile, which
are weakly affected by reaction, and linearizing close to $z=0$, this expression can be integrated over $z$ 
ans simplified to obtain (using the arguments that $\partial c/ \partial z (z\rightarrow\pm\infty)= 0$)
	\begin{equation}
\frac{{\partial {m_c}}}
{{\partial \theta }} \approx \frac{{Da}}  
{4}{\delta ^2}s_r \text{~ .}
	\label{eq:mc_Pe}
	\end{equation}
	Note that this expression is based on the same approximations as equation \eqref{eq:mass_Da_0} but it is here written in $\left\lbrace \theta, z \right\rbrace$ coordinates to account for shear.
	Recalling that $d\theta/dt=1/\delta^2$, we can rewrite equation \eqref{eq:mc_Pe} as
	\begin{equation}
	\frac{{\partial {m_c}}}
{{\partial \theta }} \approx \frac{{Da}}
{4}{\left( {\frac{{d\theta }}
{{dt}}} \right)^{ - 1}}s_r \implies \frac{{\partial {m_c}}}
{{\partial t}} \approx \frac{{Da}}
{4}\sqrt \theta   \text{~ .}
	\label{eq:red_mc_Pe}
	\end{equation}
where we have made use of the fact that during this warped time, the diffusive growth of the width is going to be proportional to $\sqrt{\theta}$. By utilizing the form of the warped time, $\theta = \left( t + Pe^2t^3/3\right) \sim Pe^2t^3/2$, we finally obtain

\begin{equation}
\frac{\partial m_c}{\partial t} \approx \frac{Da Pe}{4\sqrt{3}}t^{3/2} \implies m_c \approx \frac{DaPe}{10\sqrt{3}}t^{5/2} \text{~ .}
\label{eq:mc_Pe_final}
\end{equation}
	Hence, it is observed that the coupling of shear enhanced mixing and kinetics limitations leads to
	a strong acceleration of the effective kinetics, which is faster than all previously known regimes.
	Consistently, the mass produced is proportional to both $Da$ and $Pe$. 
	
\begin{figure}[!ht]
\includegraphics[scale=0.4]{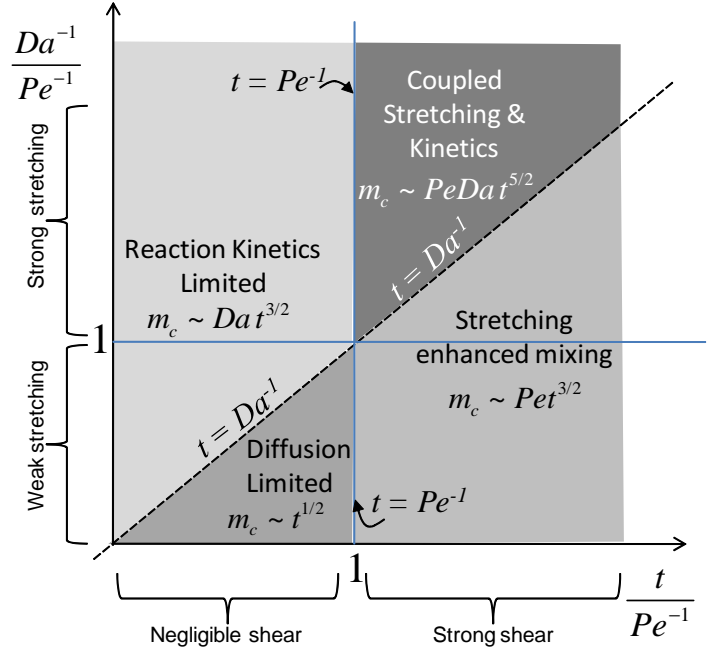}
\begin{centering}
\caption{Diagram synthesizing the different regimes predicted for the scaling of the mass of product. The $y-$axis represents the two vertical separations of weak stretching and strong stretching when $Da^{-1}/Pe^{-1} < 1$ and $>1$ respectively. The $x-$axis is also demarcated by the regimes of negligible shear and strong shear through $t/Pe^{-1} < 1$ and $>1$ respectively. }
\label{Fig:synthesis}
\end{centering}
\end{figure}

	Figure \ref{Fig:synthesis} synthesizes the different expected regimes for the scaling of the mass of produced in  mixing fronts. The new coupled stretching and kinetics regime, that shows accelerated mass production $m_c \sim Pe Da t^{5/2}$, appears to cover a significant part of the diagram. The presented analytical derivations thus provide a unified theoretical framework covering the full space of $Pe$, $Da$ and $t$ parameters.  
	
\subsection{Transition times}
\label{sec:transition}
In this section, we quantify more precisely the transition times between the different regimes discussed above. Quite naturally, in the absence of any imposed shear flow, we would expect the presence of only one transition time at $Da^{-1}$ between the kinetics-limited regime, discussed in section \ref{sec:p1}, and the diffusion-limited regime, discussed in section \ref{sec:p3}. 
The interest of the problem under consideration lies in configurations with significant imposed stretching. Depending on the relative strength of the stretching and reaction kinetics, we may have various transitions occurring as time progresses. We shall discuss these in details below.



Let us first consider the scenario of weak stretching $Da^{-1} < Pe^{-1}$. In this case, the system is expected to have two transitions. The first transition occurs between the early time kinetics-limited regime and the intermediate diffusion-limited regime, while the second transition occurs between the intermediate diffusion-limited reaction regime and the long time shear-enhanced reactive mixing regime. Towards determining the first transition time we must have an overlap of the product formed between these regions so that $Da/4 t^{3/2} = 8 t^{1/2}$ (see sections \ref{sec:p1} and \ref{sec:p3})). This yields the transition time as $t_{Da_1} = 32/Da$. Consequently, the mass of product at this transition may be found out as $m_{c,Da_1} = 45/\sqrt{Da}$. Moving on to the second transition from the diffusion-limited reaction regime (section \ref{sec:p3}) to the long time shear-mixing-limited reaction regime (section \ref{sec:p4}), we equate the masses of product for the two regimes, obtaining $8t^{1/2} = 8Pet^{3/2}$, which yields a characteristic time $t_{Pe_1} = Pe^{-1}$. In this case the mass of product at the transition is given by $m_{c,Pe_1} = 8Pe^{-1/2}$.

In the second scenario, we consider the situation with strong imposed stretching $Da^{-1} > Pe^{-1}$. In this scenario, we still expect to have two transitions. The first transition will be from the kinetics-limited regime to the intermediate coupled stretching enhanced mixing and kinetics limited regime while the second transition will be from this intermediate regime to the long time shear enhanced mixing regime with no kinetics limitation. At the first transition time, we must have an equality of the masses of product determined in the two regimes $\frac{Da}{4} t^{3/2}=\frac{Pe Da}{10\sqrt{3}}t^{5/2}$, which yields a transition time  $t_{Pe_2} = \frac{1}{Pe} \frac{5\sqrt{3}}{2}$; at this time the mass of the product is found out to be $m_{c,Pe_2} = \frac{Da}{Pe^{3/2}}\frac{1}{4}\left(\frac{5\sqrt{3}}{2} \right)^{3/2}$. Similarly, at the second transition, we have $PeDa\frac{t^{5/2}}{10\sqrt{3}} = 8 Pe t^{3/2}$. The resulting characteristic transition time is $t_{Da_2} = \frac{1}{Da} 80\sqrt{3}$, corresponding to a mass of product at the transition given by $m_{c,Da_2}  = 8 \frac{Pe}{Da^{3/2}}(80\sqrt{3})^{3/2}$.

\section{Temporal dynamics of the reactive mixing scale}
\label{sec:widths}

We now consider the temporal behavior of the reactive mixing scale $s_r$, which characterizes the spatial extent of the zone where reactions take place (equation \eqref{eq:rxnwidth}). The mixing of two separated reactants creates a hotspot of reaction at the interface, which may be highly localized in space. This can be due to i) slow diffusion, ii) fast reactions, which implies that reactants are immediately consumed as they interpenetrate each other, and iii) compression of the interface due to shear deformation. In the following, we quantify the interplay of these different processes that determines the evolution of the reactive mixing scale.

	\subsection{Dynamics of the reactive mixing scale in the absence of shear flow}
	\label{sec:noflowwidth}
	For the scenario where there is no imposed velocity gradient, the evolution of the reactive mixing scale at sufficiently long times has been mathematically demonstrated earlier \cite{Larralde1992}. For completeness, we mention it here. It was seen in \ref{sec:p1} that the initial scaling of the reactive mixing scale, for $t< Da^{-1}$, is expected to be diffusion controlled since reactions are too slow to affect the diffusive growth of the interpenetration zone of the two reactants, resulting in 
\begin{equation}
	s_r \sim \sqrt{t}. 
\label{eq:diffusive_mixing_scale}
\end{equation}
After this first regime, for $t> Da^{-1}$, reactions can be considered to be fast compared to the observation time. Hence, the progression of the reactive mixing scale is impeded by consumption of the diffusive reactants. Therefore it evolves at a much slower rate.
The temporal scaling of the reactive mixing scale can be estimated by considering the form of the reaction rate in equation \eqref{eq:R_Airy_function}, which follows a scaling form $R=f(x/t^{1/6})$. In this case, the similarity variable is $x/t^{1/6}$ since the reaction rate can be expressed solely as a function of this variable.	Hence, the reactive mixing scale, defined by the second moment of $R$ in equation \eqref{eq:rxnwidth}, is expected to follow a scaling \cite{Larralde1992}
\begin{equation}
	s_r \sim t^{1/6}. 
\label{eq:diffusive__reactive_mixing_scale}
\end{equation}
	
	\subsection{Scaling behavior with imposed flow: strong stretching regime, $Pe^{-1}< Da^{-1}$}
	\label{sec:strongstretch}
In the scenario of strong stretching, we focus on times larger than $Pe^{-1}$ because before this characteristic time the reactive mixing scale is expected to be diffusion controlled, a case which has been discussed above. 
For times $t<Da^{-1}$, reaction can be considered to be too slow to affect the progression of the reactive mixing scale. 
Hence, we can apply the same approximation as in the above case but in the $z-\theta$ coordinates, which account for shear deformation. 
In analogy to equation \eqref{eq:diffusive_mixing_scale}, the reactive mixing scale is expected to grow as $z_r \sim \sqrt{\theta}$ (diffusion-limited growth) where $z_r$ denotes the reactive mixing scale in the $z-\theta$ coordinates (we recall that $z = x'/{\delta}$). 
Therefore, the reactive mixing scale in the $x'$ coordinate is 
\begin{equation}
s_r \sim \delta \sqrt{\theta} \text{~.} 
\end{equation}
Through the heuristic argument that the form of the width is given by the above equation, we utilize the above functional relationship to arrive at a governing equation for the width. Therefore, we differentiate this with respect to time, $t$, to obtain 
\begin{equation}
\frac{1}{s_r}\frac{\partial s_r}{\partial t} = \frac{1}{\delta}\frac{\partial \delta}{\partial t} + \frac{1}{2s_r^2}\text{~.} 
\label{eq:width_1}
\end{equation}
The above equation governs the compression and diffusion of a front for the case of strong stretching at times for $t<Da^{-1}$. For $t>Da^{-1}$, the expected scaling is the same as discussed above at long times (equation \eqref{eq:diffusive__reactive_mixing_scale}).

Note that equation \eqref{eq:width_1} is the same as for the conservative mixing scale \cite{Villermaux2012, LeBorgne2015}, which is consistent with the fact that reactions are to slow to affect the progression of the front in this regime. Qualitatively, the first term depicts the compression of the reactive mixing scale due to the axial stretching and the associated perpendicular compression. On the other hand, the second term depicts the increase in the reactive mixing scale due to diffusion broadening at times where the reaction is occurring. The balance of these two terms is reached at the mixing time \cite{Villermaux2000,Meunier2010}. Before the mixing time, compression is expected to dominate over diffusion, while the opposite situation develops after the mixing time for linear shear flows. 
%
%
The mixing time can be estimated by recalling that we must have, at the transition, the balance between the width compression (the first term on the right hand side of equation \ref{eq:width_1}) and the diffusive growth (the second term on the right hand side of equation \ref{eq:width_1}). In the compression regime, the width of the reaction zone is given by the solution of $\frac{1}{s_r}\frac{\partial s_r}{\partial t} \sim \frac{1}{\delta}\frac{d\delta}{dt} \text{~.} $ which trivially yields the solution as $s_r \sim \delta$. We may thus write $\frac{1}{\delta}\frac{d\delta}{dt} \sim \frac{1}{2\delta^2}$ (where we have utilized the fact that $s_r \sim \delta$ at the transition). Simplifying this, we obtain the mixing time as $t_m \sim Pe^{-2/3}$ \cite{Villermaux2012}.

In summary, in the compression regime, $Pe^{-1}<t<Pe^{-2/3}$, the reactive mixing scale is expected to evolve by compression as $s_r \sim \frac{s_0}{\sqrt{1+Pe^2t^2}}$. In the diffusion regime, $Pe^{-2/3}<t<Da^{-1}$, it is expected to grow diffusively as $s_r \sim \sqrt{t}$. For $t> Da^{-1}$, the progression of the reactive mixing scale is expected to be impeded by the consumption of reactants as described in the previous section and the mixing scale is expected to grow as $s_r \sim t^{1/6}$.

	\subsection{Scaling behavior with imposed flow: weak stretching, $Da^{-1}< Pe^{-1}$}
	\label{sec:weakstretch}
	We now consider the weak stretching scenario, for which the reaction time scale $t_{Da_1}$ is reached earlier than the characteristic stretching time scale, $t_{Pe_1}$. To derive the temporal scaling of the reactive mixing scale in this regime, we appeal to the long time solution of equation \eqref{eq:eq_4_1}, which can be expressed as,
\begin{equation}
g \sim f(t) Ai\left(\frac{x' Da^{1/3}}{\delta (1+Pe^2t^2)^{1/3} \sqrt{\pi \theta}^{1/3}}\right).
\end{equation}
where $f(t)$ is a function that depends only on time.	
Similarly to equation \eqref{eq:R}, this leads to a general form of the reaction rate, such as,
\begin{equation}
R=Da \frac{x'}{t^{1/2}} g \sim h(t) \frac{x'}{s_r}  Ai\left( C \frac{x'}{s_r}  \right)
\end{equation}
where $h(t)$ is a function that depends only on time, C is a constant, so that the functional form of the reaction width may be written through the same heuristic argument used above as 
\begin{equation}
s_r \sim {\delta (1+Pe^2t^2)^{1/3} \theta^{1/6}}
\label{eq:s_r_weak}
\end{equation}
	
Upon taking the log and differentiating equation \eqref{eq:s_r_weak}, we obtain 
	\begin{equation}
	\frac{1}{s_r}\frac{\partial s_r}{\partial t} = \frac{1}{\delta}\frac{\partial \delta}{\partial t} - \frac{2}{3}\frac{Pe^2t}{(1+Pe^2t^2)} + \frac{1}{6\theta}\frac{d\theta}{dt}
	\label{eq:c_d_eqn}
	\end{equation}
Thus, noting that $\frac{1}{\delta}\frac{d\delta}{dt} = \frac{Pe^2 t}{1+Pe^2t^2}$, $\frac{d\theta}{dt}=1/\delta^2$ and writing $\theta$ as a function of $s_r$ through equation \eqref{eq:s_r_weak}, we obtain the compression-diffusion equation in the case of weak stretching as
 \begin{equation}
	\frac{1}{s_r}\frac{\partial s_r}{\partial t} = \frac{1}{3\delta}\frac{\partial \delta}{\partial t} + \frac{(1+Pe^2t^2)^{2/3}}{6(t+1/3Pe^2 t^3)^{2/3} s_r^{2}}
\label{eq:cd1}
\end{equation}	
	The compressive term on the right hand side of this equation is weak as compared to the diffusive expansion (as seen from the prefactor of $\frac{1}{3}$ in equation \eqref{eq:cd1}). At small times relative to the long time analysis, the solution of equation \ref{eq:cd1} is obtained by dropping the second term. This yields $s_r \sim \delta^{1/3}$.
	
	We must also reiterate that the derivation shown above is strictly speaking true for the long time behaviour (since we have appealed to the long time Airy function behavior for the concentration perturbation). This leads asymptotically to the $t^{1/6}$ behavior. Moreover, we shall see through the numerical results that at early times, the reaction width is completely governed by the compression term $\frac{1}{\delta}\frac{d\delta}{dt}$ seen in equation 
\ref{eq:width_1}. Essentially it means that the first term appearing through the long time analysis scenario of the compression-diffusion equation \ref{eq:cd1} is not going to play any role in determining the reaction mixing scale. Rather, in the initial moments, it is simply governed by the differential equation 
\begin{equation}
\frac{1}{s_r}\frac{\partial s_r}{\partial t} = \frac{1}{\delta}\frac{\partial \delta}{\partial t} 
\label{eq:just_compress}
\end{equation}
	
	For completeness, we still attempt to evaluate the mixing time for this scenario for which $\frac{1}{3\delta}\frac{\partial \delta}{\partial t} \sim \frac{(1+Pe^2t^2)^{2/3}}{6\theta^{2/3}s_r^{2}}$. Upon substituting the form of $s_r \sim \delta^{1/3}$ in the compression regime and equating these two terms we obtain $t_m \sim Pe^{-2/3}$, which is the same as for the strong stretching regime.
	
	The expressions derived here provide governing equations for the reactive mixing scale that allow quantifying the spatial localization of chemically reactive zone across mixing fronts. While equation \eqref{eq:width_1} is the same as for the conservative mixing scale, equation \eqref{eq:cd1} is different, which quantifies the limitation of the growth of the mixing scale by the consumption of reactant through chemical reactions. Regardless of the regimes described in section \ref{sec:weakstretch}) and \ref{sec:strongstretch}, the mixing time is always equal to $Pe^{-2/3}$, which is the same as for conservative mixing.
	
\section{Numerical simulation of equations in the Lagrangian frame}
\label{sec:discrete}
In order to test the analytical expressions derived for the mass of product formed and the reactive mixing scale, we solve the set of equations \eqref{eq:warpedeq} by means of a Chebyshev spectral collocation method \cite{Trefethen2000}. This methodology has been employed in recent times to address a large assortment of nonlinear PDEs. Towards its implementation, we have to note that the domain is rescaled by $L$, where $L$ is chosen sufficiently large over the desired observation time so as to prevent the boundary effects to affect the system evolution for the time span of interest. We denote the rescaled spatial variable as  $\hat z = z/L$. The governing equations are subjected to the initial condition for $a$ as $a\left( {\hat z,0} \right) = 1\,\,\,\forall \hat z \in \left( {0,1} \right)$ and $0$ otherwise. The initial condition for $b$ is given by $b(\hat z, 0) = 1 - a(\hat z, 0)$ while for the species $c$ we have the initial condition as $c(\hat z,0) = 0$ everywhere. The boundary conditions are $a(1,\theta) = 0$, $a(-1,\theta)=1$, $b(1,\theta) = 1$, $b(-1,\theta) = 0$ and $c(\pm 1,\theta) = 0$. The method involves discretizing the 1D domain into the {\it{Gauss-Lobatto}} discretized grids represented by ${\hat z_i} = \cos \left( i\pi/N\right)$, where $N$ represents the total number of intervals chosen for the solution. We make use of the Chebyshev differentiation matrices to represent the spatial derivatives. In a time discrete form (with the superscript $k$ denoting the $k^{th}$ time level), the discretized equation for the species $a$ may be represented as 
\begin{equation}
\frac{{{a^{k + 1}} - {a^k}}}
{{\Delta \theta }} = {\left[ {{a_{xx}}} \right]^{k + 1}} - Da\,{a^k}{b^k}{\left( {{\delta ^k}} \right)^2}
\label{eq:discrete1}
\end{equation}
We may obtain the time-discrete equations for the evolution of species $b$ and $c$ along similar lines. Briefly speaking, the governing equations for the three species are expanded in the form of $n$ Chebyshev polynomials of the first kind over the discretized domain $\hat z$ as
\begin{equation}
a\left( {\hat z} \right) \approx \sum\limits_{l = 0}^n {{A_l}{T_l}\left( {\hat z} \right)} 
\label{eq:cheb}
\end{equation}
We may write the discretized form of the reaction-diffusion equations as
\begin{eqnarray}
\begin{gathered}
  a_i^{k + 1} + \Delta \theta \left[ {{a_{xx}}} \right]_i^{k + 1} = a_i^k - Da\,a_i^kb_i^k\frac{1}
{{ {1 + P{e^2}{{\left( {{t^{k + 1}}} \right)}^2}} }} \hfill \\
  b_i^{k + 1} + \Delta \theta \left[ {{b_{xx}}} \right]_i^{k + 1} = b_i^k - Da\,a_i^kb_i^k\frac{1}
{{ {1 + P{e^2}{{\left( {{t^{k + 1}}} \right)}^2}} }} \hfill \\
  c_i^{k + 1} + \Delta \theta \left[ {{c_{xx}}} \right]_i^{k + 1} = c_i^k + Da\,a_i^kb_i^k\frac{1}
{{{1 + P{e^2}{{\left( {{t^{k + 1}}} \right)}^2}} }} \hfill \\ 
\end{gathered} 
\label{eq:alleqs}
\end{eqnarray}
where the subscript $i$ represents the value of the variable at the $\i-th$ node while the superscript $k$ denotes the $k-th$ time step. The initial and boundary conditions for equation \ref{eq:alleqs} may be written as
\begin{equation}
\begin{gathered}
  a_i^0 = 1\,{\text{ }}\forall x > 0,b_i^0 = 1 - a_i^0,c_i^0 = 0 \hfill \\
  a_N^k = 1,\,a_0^k = 0,\,b_N^k = 1,\,b_0^k = 0,c_N^k = 0,\,c_0^k = 0 \hfill  \text{~ .} 
\end{gathered} 
\label{eq:bcsdisc}
\end{equation}
The spatial derivatives are represented by means of the Chebyshev differentiation matrices, $\mathcal{D}$, with $(\mathcal{D}\;a)$ representing the first derivative of the vector $a$, $(\mathcal{D}^2\;a)$ representing the second derivative of the vector $a$ and so on. The boundary conditions of the discretized domains are incorporated by altering the first and last rows of the pertinent differentiation matrix, respectively. 
The MATLAB files for obtaining the spatiotemporal evolution of the concentration profiles and other derived parameters can be made available upon request. In the appendix we have demonstrated the validity of the numerical methodology presented here against full 2D simulations performed in the finite element framework of COMSOL Multiphysics. It may be seen from the figures that the simulations are in excellent agreement. The methodology presented here requires  modest computational resources in comparison to the full 2D simulations \cite{Trefethen2000}. In the following, we use this $1D$ numerical model of diffusion, compression and reaction in the direction transverse to the front to explore the space of $Pe$ and $Da$ and validate the analytical expressions derived in the previous section.

\section{Discussion}
\label{sec:results}

In this section we shall attempt to shed light on the temporal scaling of the mass produced and trends observed for the transition times among the different regimes as obtained from the theoretical discussion in this work, and compare it against the numerical results obtained by means of the 1D Chebyshev spectral collocation method.
We shall first focus on the temporal evolution of the mass of product formed for the cases of no imposed flow and imposed flow (with the two subcases of weak stretching and strong stretching). 
We also attempt to unravel the underlying universality of the temporal evolution of the mass of the product formed, by rescaling the mass as well as the time with the suitable transition times obtained in our analysis in section \ref{sec:transition}. We then proceed to study the temporal evolution of the reactive mixing scale for the cases of no shear and shear while attempting to quantify the observed temporal evolutions by means of the compression-diffusion equation proposed in section \ref{sec:widths}.

	\subsection{Temporal evolution of the mass of product}

	\subsubsection{No shear scenario}
	
\begin{figure}[!ht]
\begin{centering}
\includegraphics[scale=0.6]{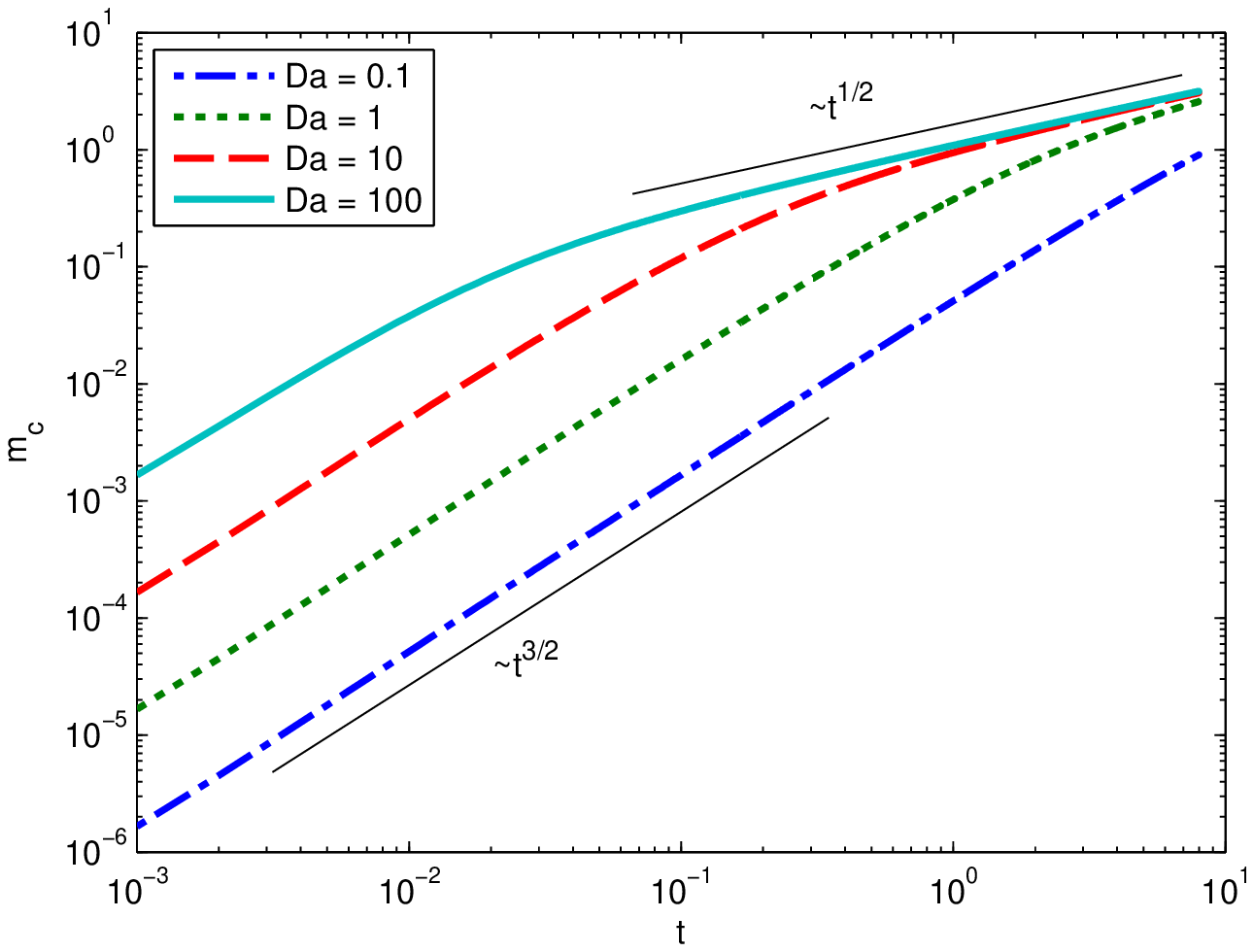}
\caption{Temporal evolution of the mass of product formed for $Da = 0.1, 1, 10, 100$ in the case where $Pe = 0$, as obtained from the Chebyshev spectral method.}
\label{Fig:mvst_Pe_0}
\includegraphics[scale=0.6]{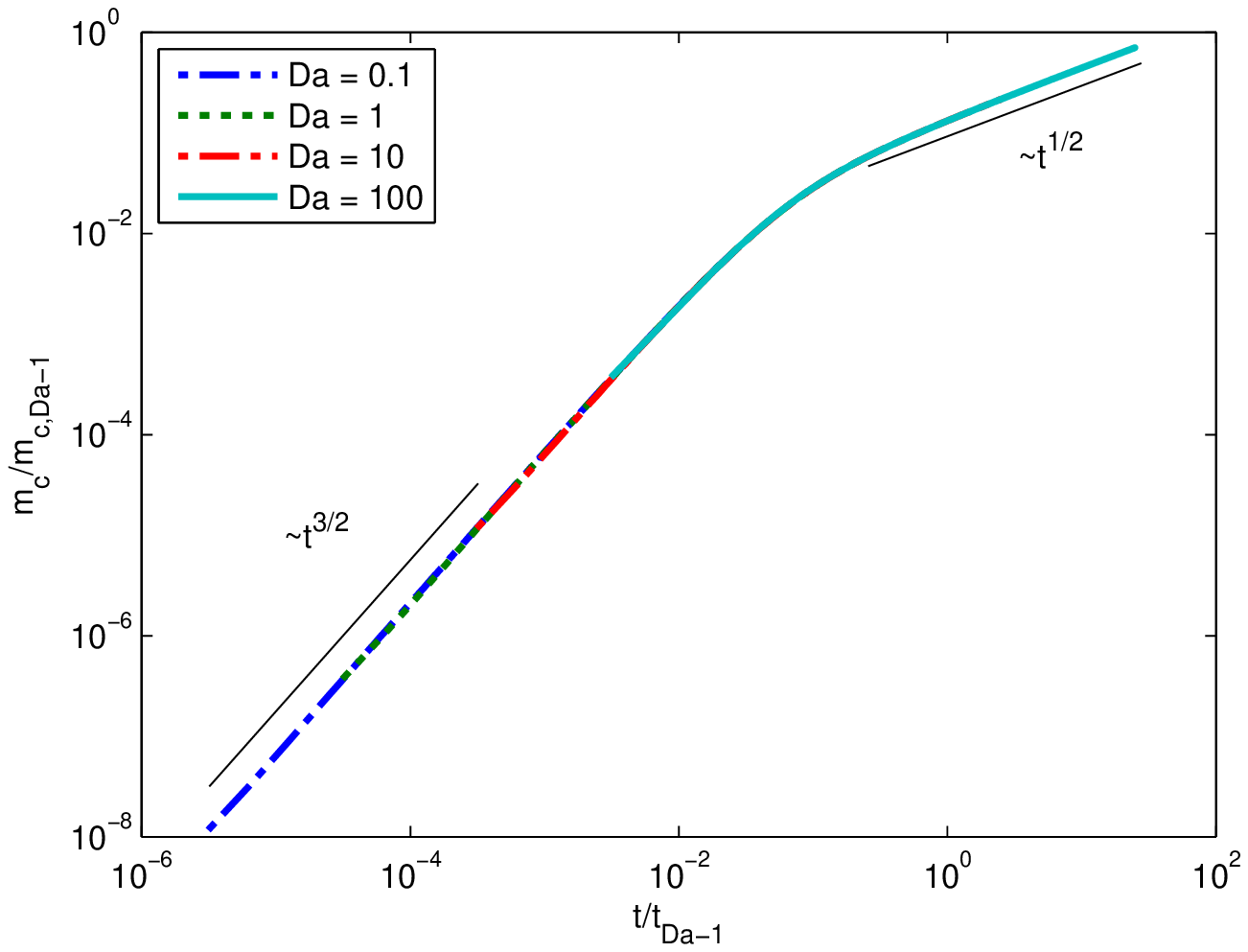}
\caption{Rescaled temporal evolution of the mass of product formed for $Da = 0.1, 1, 10, 100$ for the case where there is no externally imposed flow. Towards rescaling the aforementioned variables, the time has been rescaled as $t/(32/Da)$ while the mass has been rescaled as $m_c/(45/\sqrt{Da})$.}
\label{Fig:mvst_scaled_Pe_0}
\end{centering}
\end{figure}	

	For completeness, we first depict the temporal evolution of the mass of product for the case where there is no imposed flow. In figure \ref{Fig:mvst_Pe_0} we depict the temporal variation of the mass of the product formed for different Damk\"ohler numbers ($Da = $0.1, 1, 10, 100). 
	We observe from figure \ref{Fig:mvst_Pe_0} that the initial scaling obtained from the numerical simulation does corroborate excellently with the theoretical predictions made in section \ref{sec:p1}; in the initial reaction-diffusion regime (reaction kinetics dominated regime), the mass scales as $m_c \sim Da t^{3/2}$, as predicted analytically.
	Essentially we observe that as $Da$ increases, there is a concomitant increase in the product mass produced. Beyond the transition time $t_{Da_1}= 32/Da$ the mass scaling is no more limited by reaction kinetics and evolves as per the scaling $m_c \sim t^{1/2}$, corresponding to the diffusion-limited regime, as theoretically obtained in section \ref{sec:p3}. In accordance to the theoretical predictions, it is seen from figure \ref{Fig:mvst_Pe_0} that beyond the time $t_{Da_1}$, $m_c$ becomes independent of the Damk{\"o}hler number. Clearly, this underpins the fact that beyond the transition time, the evolution is limited by diffusion and not dictated by the reaction kinetics term. 
In figure \ref{Fig:mvst_scaled_Pe_0}, we depict the temporal evolution of the rescaled mass of the product $m_c/m_{c,Da_1}$ as a function of the rescaled time $t/t_{Da_1}$. The curves depicted in figure \ref{Fig:mvst_Pe_0} collapse on top of each other, thus confirming the validity of the theoretical predictions of the rescaling of the transition time and the mass of the product formed at that particular transition time. 

\subsubsection{Weak stretching scenario}

\begin{figure}
\includegraphics[scale=0.6]{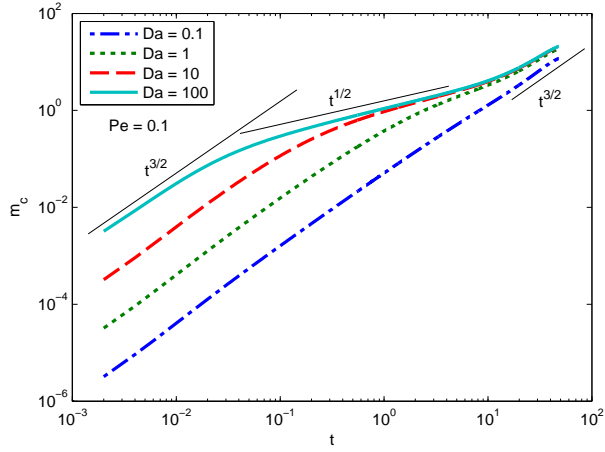}
\begin{centering}
\caption{Time evolution of the product mass for $Pe = 0.1$ for $Da = 0.1, 1, 10, 100$. The three distinct regimes, reaction controlled, diffusion controlled and stretching controlled, are observed when the two transition times are separated (i.e. $Pe <  Da$). As $Pe$ becomes comparable to $Da$, the intermediate regime of diffusion-limited growth is not observed.}
\label{Fig:mvst_Pe0.1}
\end{centering}
\end{figure}

\begin{figure}[!th]
\includegraphics[scale=0.6]{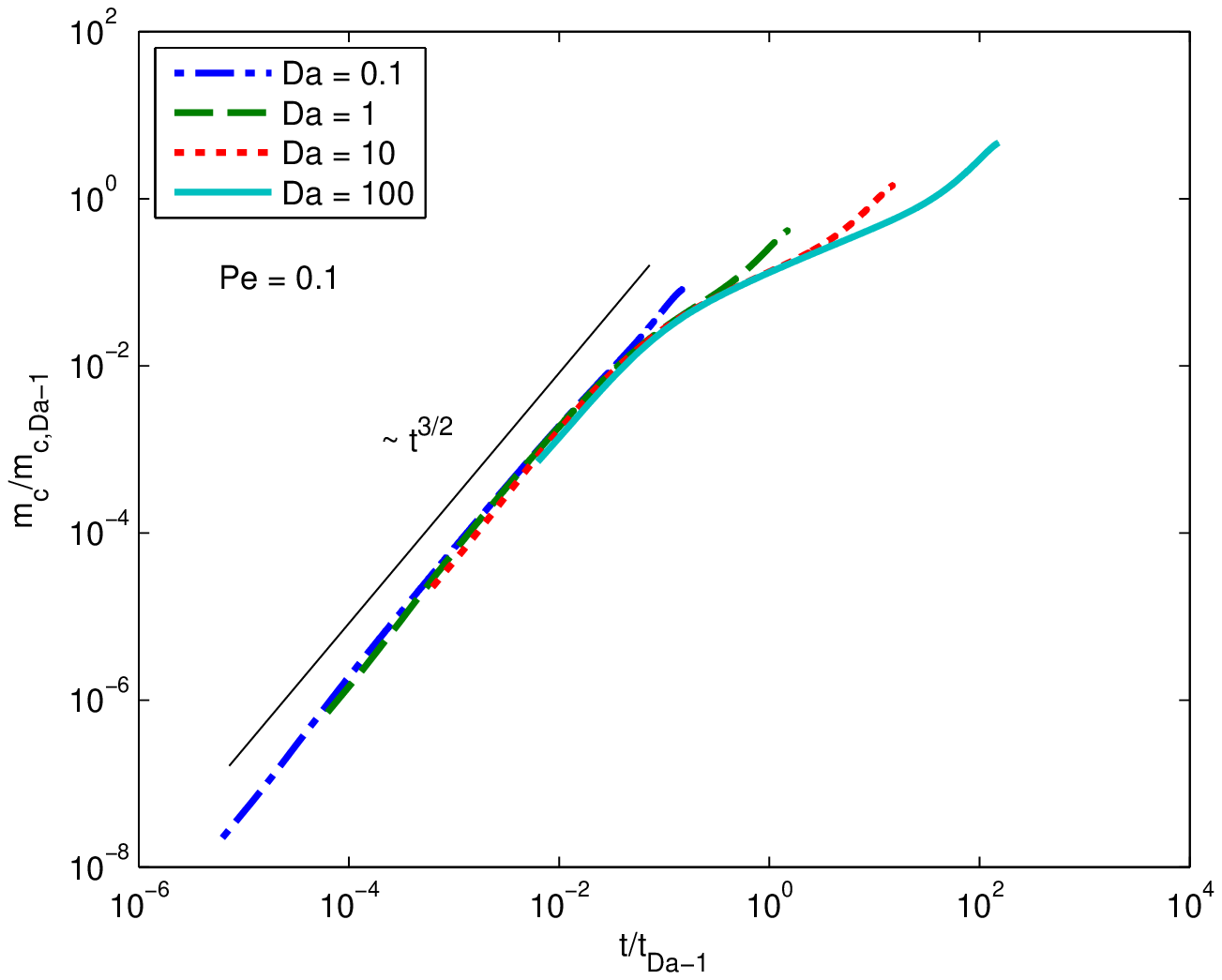}
\begin{centering}
\caption{Scaled mass of product vs time $m_c/m_{c,Da_1}$ as a function of $t/t_{Da_1}$, for $Pe = 0.1$ and $Da = 0.1, 1, 10, 100$ at the first transition time for weak stretching, $t_{Da_1}$.}
\label{Fig:mvst_Pen0_weakstretch_T1}
\includegraphics[scale=0.6]{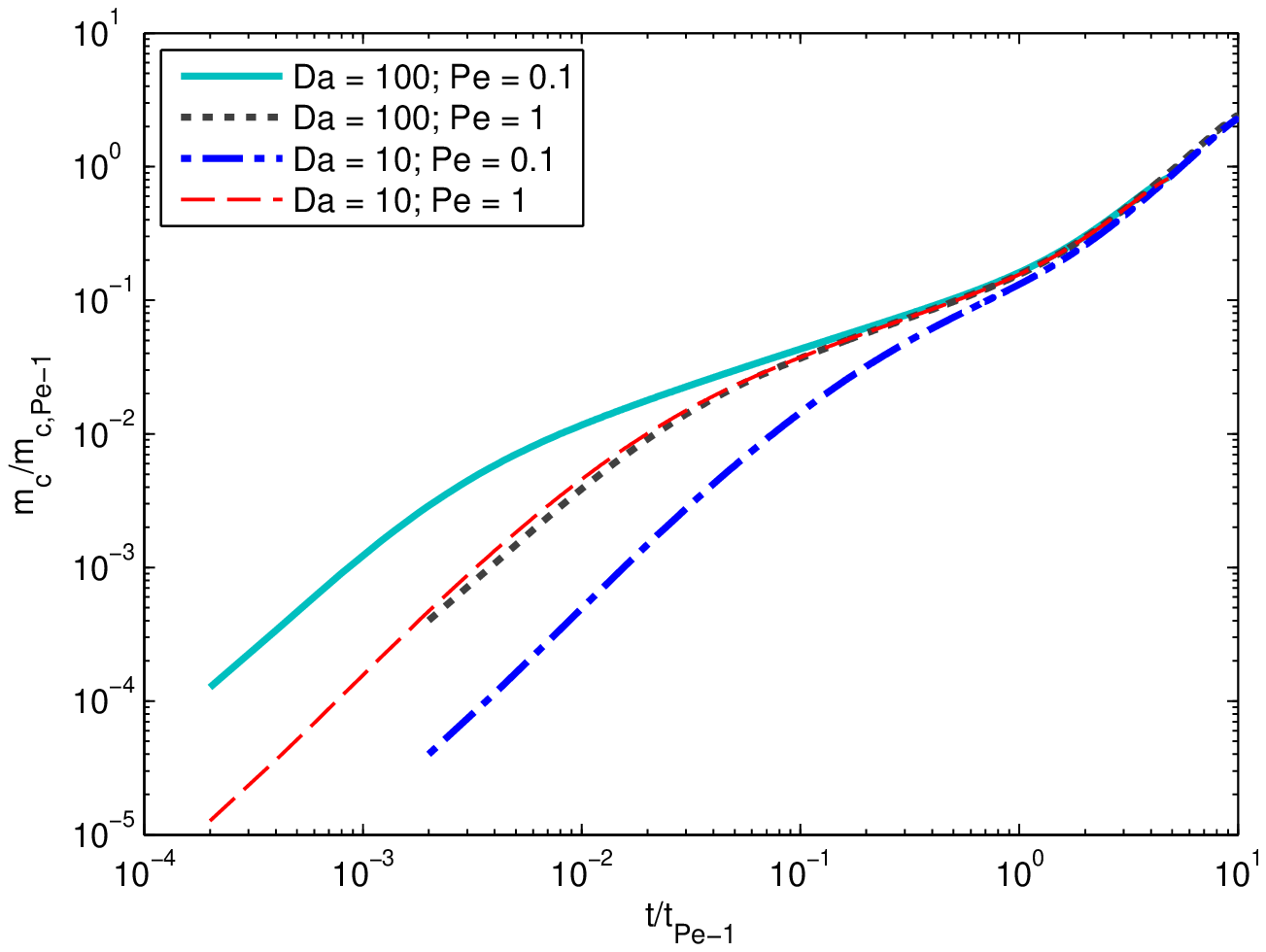}
\caption{Scaled mass of product vs time, for $Pe=0.1$ and $Da = 0.1, 1, 10, 100$ at the second transition time for weak stretching, $t_{Pe-1}$. The inset depicts the same rescaled plots for $Pe = 1$ and $Da = 1, 10, 100$.}
\label{Fig:mvst_Pen0_weakstretch_T2}
\end{centering}
\end{figure}

We now proceed to analyze the weak stretching scenario $Da^{-1} < Pe^{-1}$. In figure \ref{Fig:mvst_Pe0.1} we depict the temporal evolution of the mass of the product formed for $Pe = 0.1$ and $Da = 0.1, 1, 10, 100$. Initially all the curves increase as per the scaling $m_c \sim t^{3/2}$, indicating the kinetics-limited regime. After this initial scaling, depending on the magnitude of $Da$, a transition from the $m_c \sim t^{3/2}$ scaling to the $m_c \sim t^{1/2}$ scaling is observed. The dependence of the transition time on $Da$ is consistent with the predicted relationship $t\sim Da^{-1}$. Then at around $t\sim Pe^{-1}$ all the curves merge together to follow a mass scaling in the form $t^{3/2}$, which is consistent with the expected shear-enhanced reactive mixing regime. For the case $Da = Pe= 0.1$, the intermediate regime does not exist and the system transits directly from the kinetics limited regime to the long-time shear-enhanced reactive mixing regime (wherein both the regimes have a temporal scaling of $m_c \sim t^{3/2}$). 
Besides this, we also notice that, in accordance to the theoretical prediction, the long time behavior of $m_c$ is independent of $Da$. 
Upon rescaling the curves of figure \ref{Fig:mvst_Pe0.1} by the first transition time $t_{Da_1} = 32/Da$ for the time axis, and the mass of product at this time $m_{c,Da_1} \approx 45/\sqrt{Da}$, they collapse into each other (figure \ref{Fig:mvst_Pen0_weakstretch_T1}) for the first two regimes (i.e. for $t<t_{Pe_1}$). This confirms the validity of the predicted scalings for the first transition time $t_{Da_1}$. 
In figure \ref{Fig:mvst_Pen0_weakstretch_T2} we observe that the curves for different $Pe$ (and quite obviously different $Da$ as well) collapse onto each other for the later two regimes (i.e. for $t>t_{Da_1}$) when we rescale time for the the curves in figure \ref{Fig:mvst_Pe0.1} by the second transition time $t_{Pe_1} \approx Pe^{-1}$ and the mass of the product formed by $m_c \approx 8Pe^{-1/2}$.

We remark that all the observations are in complete agreement with the theoretical predictions presented in sections \ref{sec:p1}, \ref{sec:p3}, and \ref{sec:p4}. The regime $Da^{-1}<Pe^{-1}$ thus essentially decouples the effect of stretching, active for $t>Pe^{-1}$, from the effect of kinetics limitations, which are significant for $t<Da^{-1}$. In what follows, we shall consider the case where the fluid stretching is coupled to the chemical reactions, i.e. the strong stretching scenario where $Pe^{-1}<Da^{-1}$.
 
\subsubsection{Strong stretching scenario}

\begin{figure}
\begin{centering}
\includegraphics[scale=0.6]{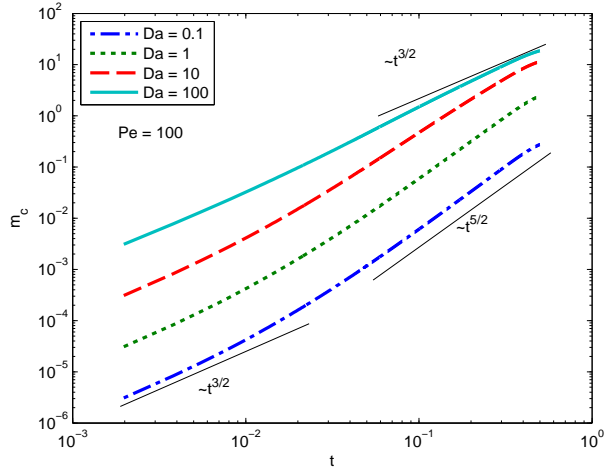}
\caption{Time evolution of the product mass for $Pe = 100$ and $Da = 0.1, 1, 10, 100$. As opposed to the case considered in figure \ref{Fig:mvst_Pe0.1}, we consider the effect of strong stretching, thus allowing the system to undergo the transition into the stretching controlled regime at early times.}
\label{Fig:mvst_Pe100}
\end{centering}
\end{figure}

In figure \ref{Fig:mvst_Pe100} we investigate the strong stretching scenario $Pe^{-1}<Da^{-1}$. We consider the temporal evolution of $m_c$ for the cases where $Pe = 100$ and $Da = 0.1, 1, 10, 100$. For $t<Pe^{-1}$, we observe that the evolution of $m_c$ scales as $m_c \sim t^{3/2}$. After this time, stretching becomes significant while reaction kinetics is still a limiting factor since $t<Da^{-1}$.  Hence, in this intermediate regime we observe the predicted accelerated scaling $m_c \sim t^{5/2}$, induced by the coupling between kinetics limitation and stretching. This first transition is confirmed by rescaling time by $t_{Pe_2} = 5\sqrt{3}Pe^{-1}/2$ and mass by $m_{c, Pe_2} = (5\sqrt{3}/2)^{3/2}/Pe^{3/2}$, which collapses the curves together (except for the case of $Da = 100$, which does not meet the requirement of strong stretching owing to the fact that we have $Da \sim Pe$). The second transition from $m_c \sim t^{5/2}$ regime to $m_c \sim t^{3/2}$ regime is not visible on a single simulation due to very long time of simulation required. This transition is however visible when collapsing the different simulations together through rescaling time by $t_{Da_2} \approx 80\sqrt{3}/Da$ and mass by $m_{c,Da_2} \approx 8(80\sqrt{3})^{3/2}Pe/Da^{3/2}$ (figure \ref{Fig:mvst_Pen0_strongstretch_T2}). 

\begin{figure}[!th]
\begin{centering}
\includegraphics[scale=0.6]{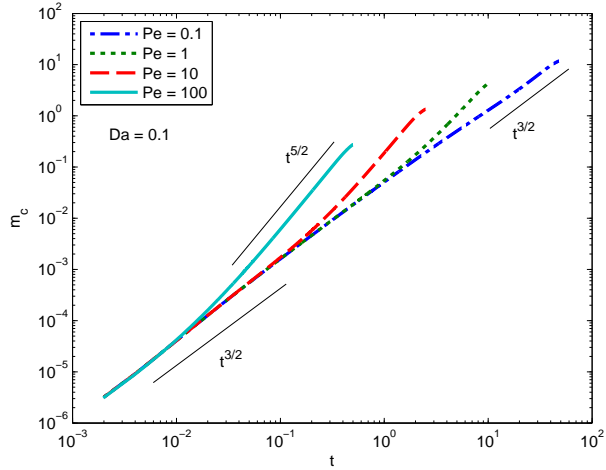}
\caption{Time evolution of the product mass for $Da = 0.1$ and $Pe = 100, 10, 1, 0.1$.}
\label{Fig:mvst_diff_Da_Pe_100}
\end{centering}
\end{figure}

Note that the intermediate scaling is not observed for $Pe=Da=100$, since the system transitions directly from the early time kinetics-limited regime with no significant stretching to the stretching enhanced reactive mixing regime with no kinetics limitation, which both scale as $m_c \sim t^{3/2}$. 
The agreement with theoretical derivations is further confirmed by the series of simulations performed for $Da = 0.1$ and $Pe = 0.1, 1, 10, 100$ (figure \ref{Fig:mvst_diff_Da_Pe_100}). The transition $t_{Pe_2}$ is readily seen in this case  when considering the trends of the transition from the $m_c \sim t^{3/2}$ regime to the $m_c \sim t^{5/2}$ regime as $Pe$ is increased. For the situation for which $Da \sim Pe$, we see again that the region in which there is a coupling of the reaction kinetics and stretching vanishes as expected. Figure  \ref{Fig:mvst_Pen0_strongstretch_T1} clearly depicts that when the time and the mass of the product are rescaled by $t_{Pe_2}$ and $m_{c,Pe_2}$ respectively, we observe that the curves in figure \ref{Fig:mvst_Pe100} collapse onto each other for the first two regimes (i.e. for $t<Da^{-1}$. On the other hand figure  \ref{Fig:mvst_Pen0_strongstretch_T2} clearly depicts that when the time and the mass of the product are rescaled by $t_{Da_2}$ and $m_{c,Da_2}$, we obtain a good collapse of the curves depicted in figure \ref{Fig:mvst_Pe100} for the other later two regimes (i.e. for $t > Pe^{-1}$).

\begin{figure}[!th]
\begin{centering}
\includegraphics[scale=0.6]{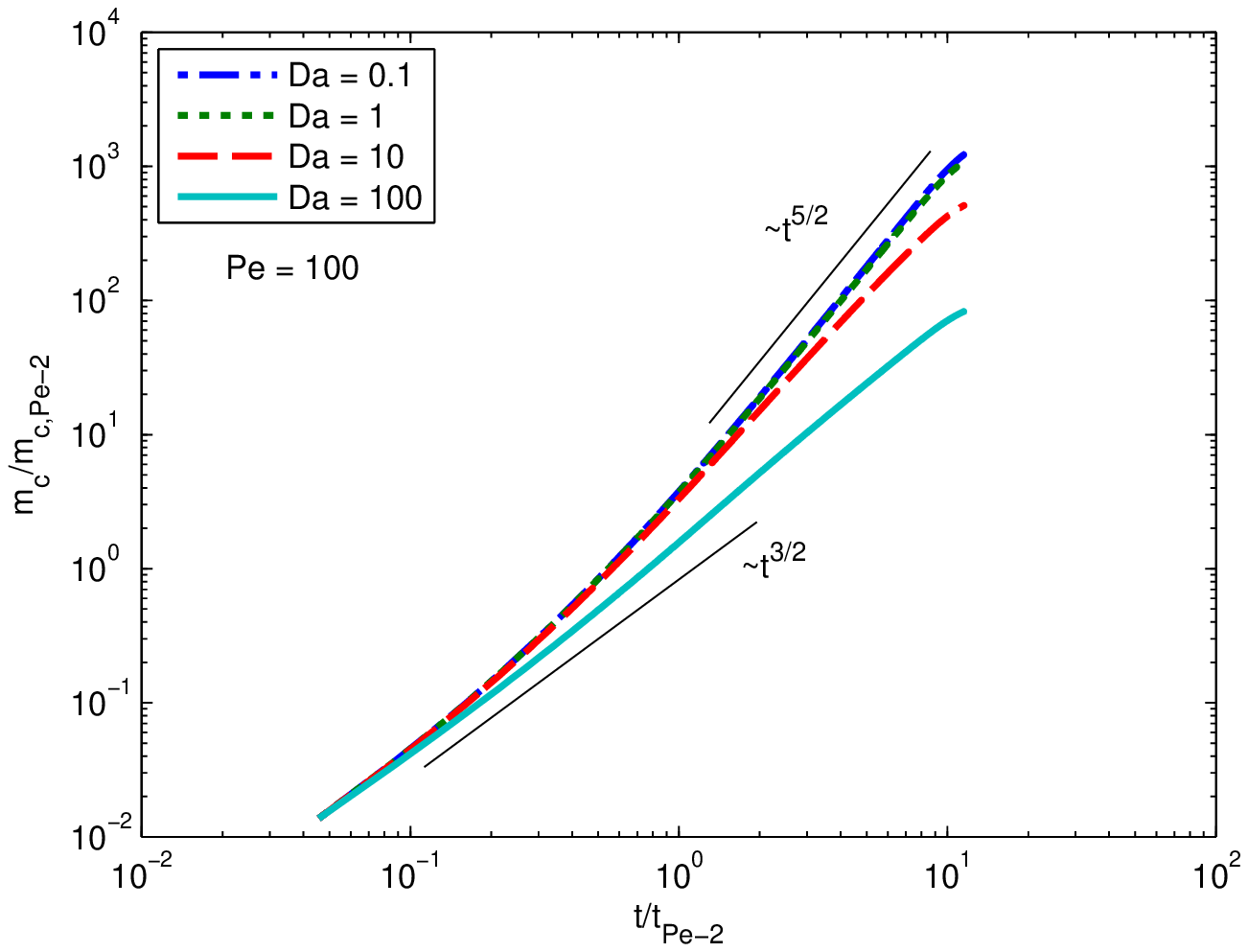}
\caption{Rescaled mass of product vs time, for $Pe = 100$ (strong stretching) and  $Da = 0.1, 1, 10, 100$ at the first transition time, $t_{Pe-2}$.}
\label{Fig:mvst_Pen0_strongstretch_T1}
\includegraphics[scale=0.6]{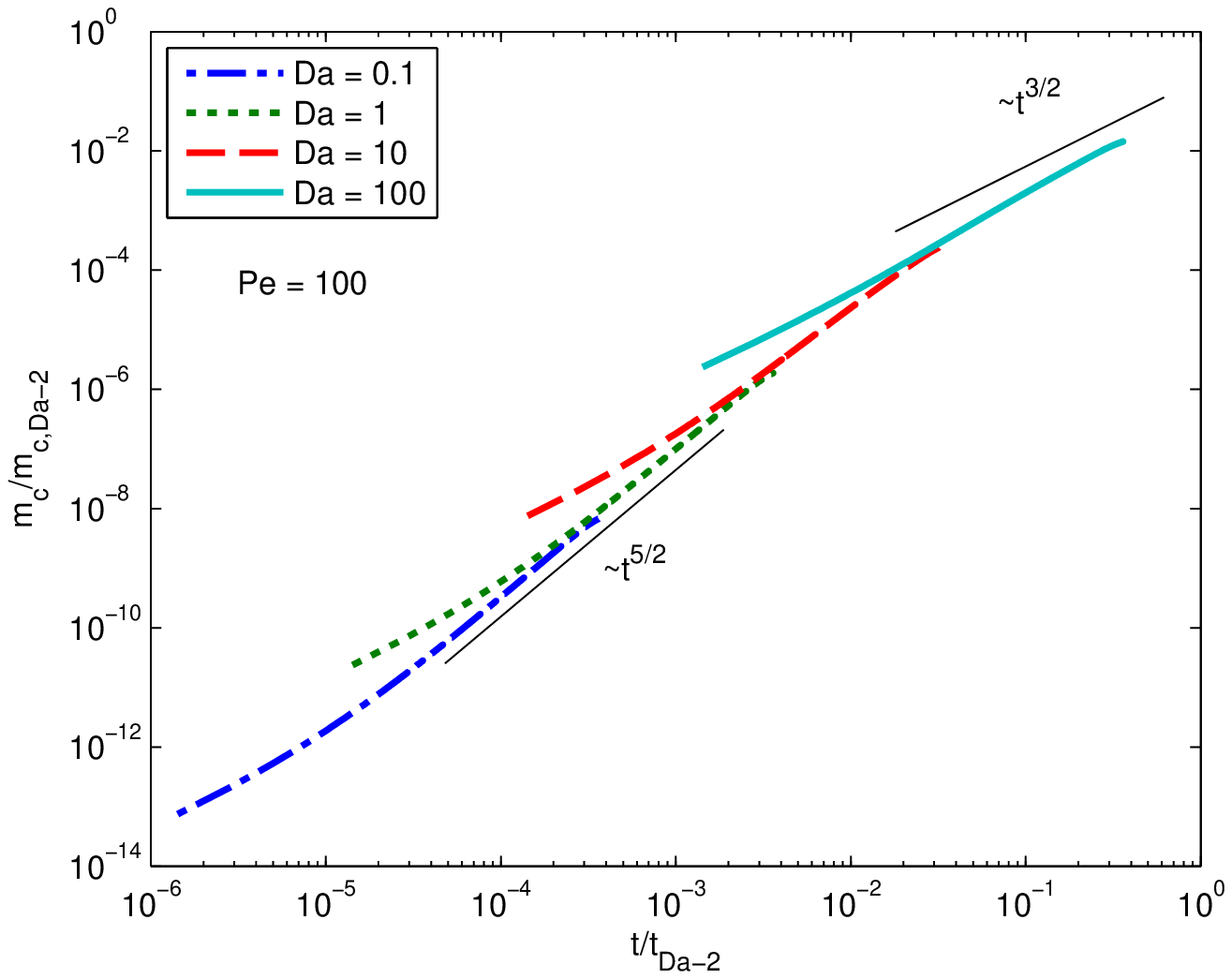}
\caption{Rescaled mass of product vs time, for $Pe = 100$ (strong stretching) and $Da = 0.1, 1, 10, 100$ at the second transition time, $t_{Da-2}$.}
\label{Fig:mvst_Pen0_strongstretch_T2}
\end{centering}
\end{figure}

	\subsection{Temporal evolution of the reactive mixing scale}
	
\begin{figure}
\includegraphics[scale=0.6]{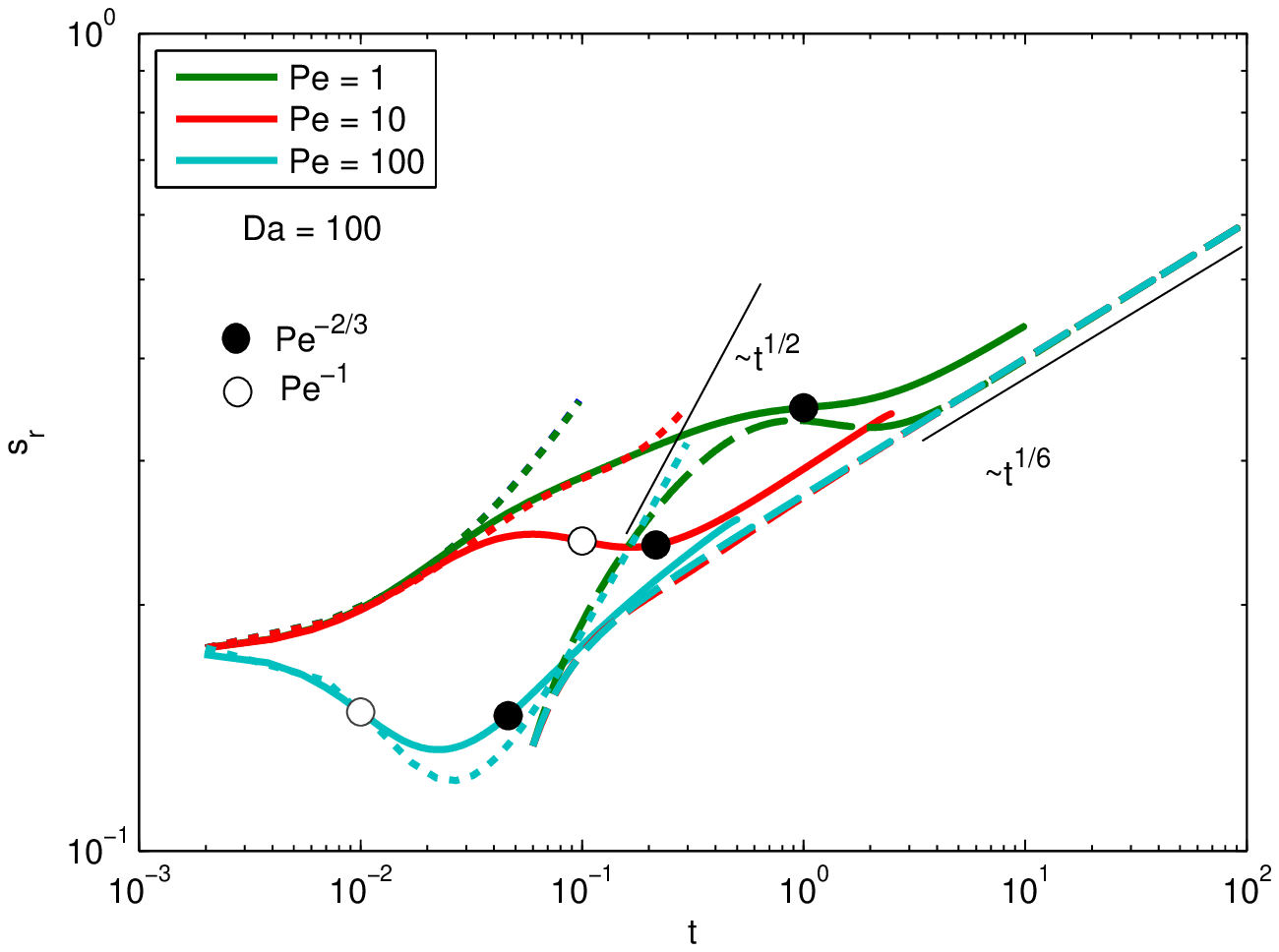}
\begin{centering}
\caption{Temporal evolution of the reactive mixing scale as a function of time for $Da = 100$ and $Pe = 1, 10, 100$. The black filled markers correspond to the mixing time $t \sim Pe^{-2/3}$ while the white filled marker correspond to the time $Pe^{-1}$ for the cases $Pe = 10$ and $100$.
The solid lines represent the numerical solution obtained using the 1D Chebyshev spectral collocation method, the dotted lines represent the solution obtained from the compression-diffusion equation in section \ref{sec:strongstretch}, and the dashed curves represent the curves obtained using the compression-diffusion equation in section \ref{sec:weakstretch}.
}
\label{Fig:stretching1}
\end{centering}
\end{figure}	
In figure \ref{Fig:stretching1} we depict the evolution of the reactive mixing scale as a function of time for configurations of weak stretching with $Da = 100$ and $Pe = 0.1, 1, 10, 100$. 
The compression-diffusion equation which valid in the early time regimes (equation \eqref{eq:width_1}, derived in section~\ref{sec:strongstretch}) is able to predict the early time behavior for the curves to a good approximation. 

As explained in section~\ref{sec:strongstretch}, at small times the reactive mixing scale grows at a rate which is limited by diffusion; the effect of stretching has not yet come into the picture. In this regime, owing to the reaction taking place between the two reactants, there is a diffusion controlled growth of the reactive mixing scale. This occurs until the time $t\sim Pe^{-1}$ at which a compression occurs in the system. This is most prominently seen in the case of $Pe = 100$ and $Pe = 10$. In-between the stretching time $Pe^{-1}$ and  and the mixing time $Pe^{-2/3}$ we observe that the impact of the stretching is to compress the reactive mixing scale. This compression occurs until the mixing time $t_m \sim Pe^{-2/3}$, beyond which, for the case of relatively weak stretching, we observe a growth of the reactive mixing scale as $s_r \sim t^{1/6}$. Therefore, the initial time behavior is well captured by equation \eqref{eq:width_1} while the late time behavior is well captured by equation \eqref{eq:cd1}. 

\begin{figure}[!th]
\begin{centering}
\includegraphics[scale=0.6]{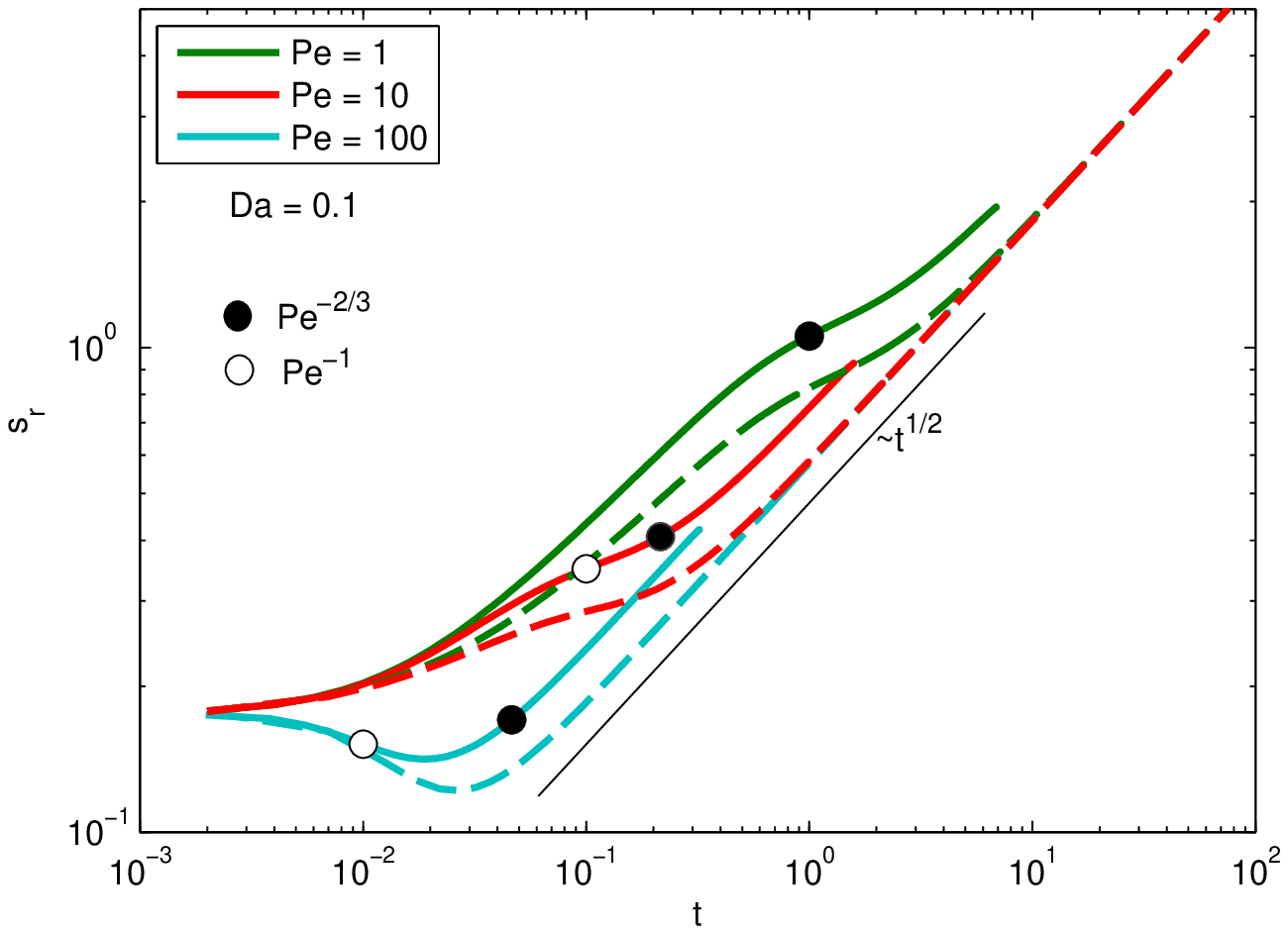}
\caption{Temporal evolution of the reactive mixing scale for a case for which $Pe$ is larger than $Da$. We have chosen $Da = 0.1$ and depicted the reactive mixing scales for $Pe = 1, 10, 100$. The black filled markers corresponds to the mixing time $t_m = Pe^{-2/3}$ while the white filled marker corresponds to the time $Pe^{-1}$ for $Pe = 10$ and $100$. The solid lines represent the solution from the 1D Chebyshev spectral collocation method while the dashed lines represent the prediction from the compression diffusion equation \ref{eq:width_1}}
\label{Fig:stretching2}
\end{centering}
\end{figure}	
In figure \ref{Fig:stretching2} we depict the temporal evolution of the reactive mixing scale, $s_r$ as a function of time for configurations of strong stretching as manifested through a low Damk\"ohler ($Da = 0.1$) and a P{\'e}clet that is larger than the Damk{\"o}hler ($Pe = 0.1, 1, 10, 100$).  
We observe a good agreement with the compression-diffusion equation elucidated in section \ref{sec:strongstretch}.
There is an initial diffusion controlled growth of the reactive mixing scale until the time $Pe^{-1}$. After this time we observe a compression (the reactive mixing scale decreases), followed by a transition at $t \sim Pe^{-2/3}$ to the $s_r \sim t^{1/2}$ regime. 

\begin{figure}
\begin{centering}
\includegraphics[scale=0.6]{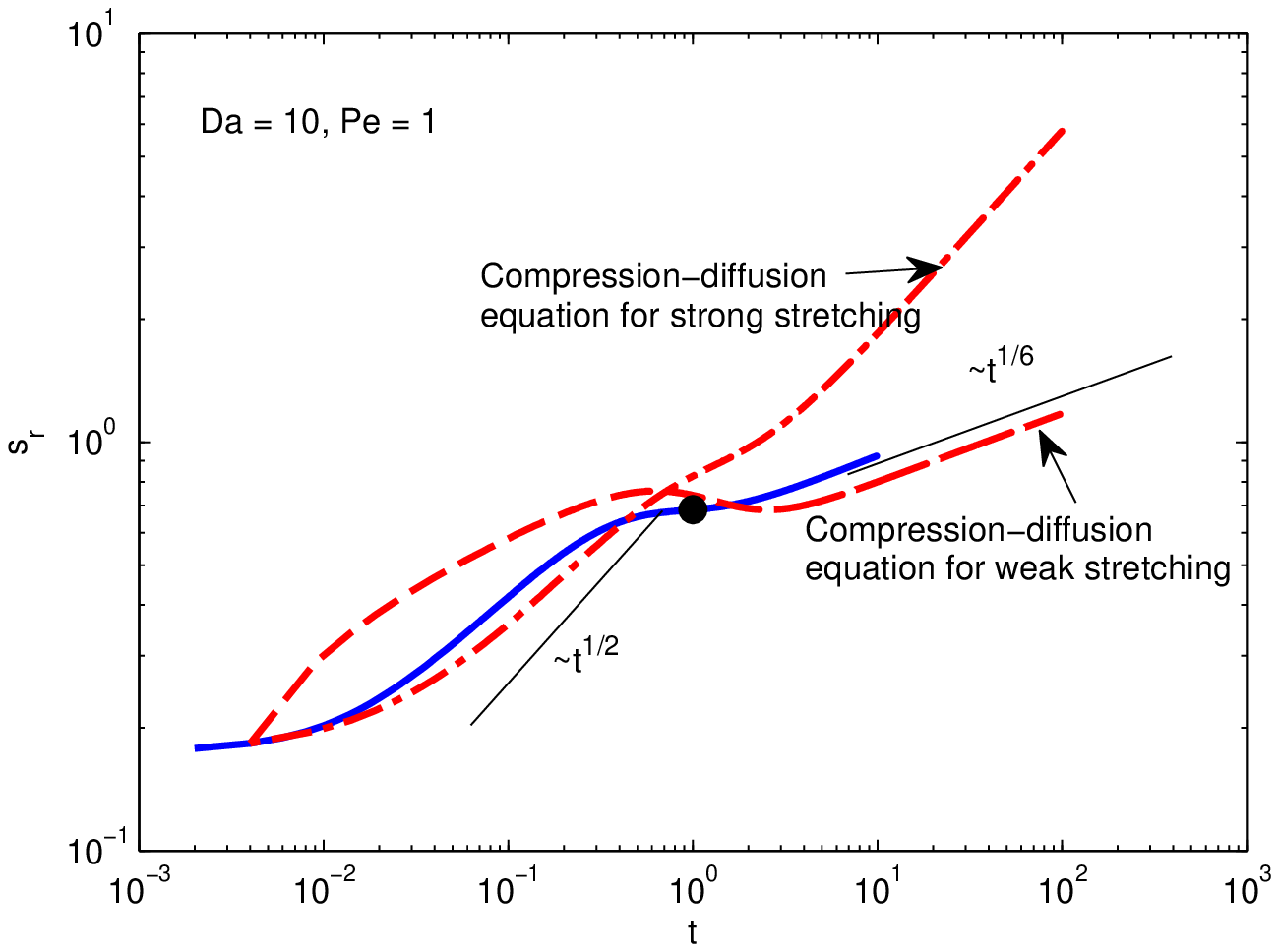}
\caption{Temporal evolution of the reactive mixing scale as a function of time for the case with $Da = 10$ and $Pe = 1$. The solid line represents the solution obtained from the 1D Chebyshev spectral collocation method. The dashed and the dash-dot-dash line represents the solution from the compression diffusion equations \ref{eq:cd1} and \ref{eq:width_1} respectively}
\label{Fig:fig13}
\end{centering}
\end{figure}
	Moving beyond the extreme cases of $Pe$ and $Da$ in figure \ref{Fig:fig13} we represent the temporal evolution of the reactive mixing scale for a situation where $Pe$ and $Da$ are moderate and not much different from each other in a hope to outline the transition from one kind of behavior to the other kind of behavior. For this case we have chosen $Da = 10$ and $Pe = 1$. The solid line represents the numerical solution while the dash-dot line represents the solution obtained from the compression-diffusion equation \eqref{eq:width_1} while the dashed line represents the solution obtained from the compression-diffusion equation \eqref{eq:cd1}. It may be clearly observed that the initial evolution of the reactive mixing scale is reasonably predicted by the compression-diffusion equation \eqref{eq:width_1} while beyond the overlap of the two curves, the curve is depicted in a better fashion by \eqref{eq:cd1}. The crossover in behavior occurs at the the mixing time scale, which in this case is close to $1$. 
	
\section{Conclusions}

We have comprehensibly studied the dynamics of reacting fronts in the presence of shear flow, which is an important fluid deformation process expected to dictate the evolution of reactive mixing fronts from the pore scale to the catchment scale. The existence of velocity gradients promotes mixing by elongating the interface available for diffusive mass transfer and increasing concentration gradients. As discussed in this study, the coupling of this phenomenon with a broad range of characteristic reaction time scales is non-trivial. We have thus quantified the impact of the Damk{\"o}hler and P{\'e}clet numbers on the effective reaction kinetics of mixing front subject to shear deformation. Different regimes have been identified theoretically in this temporal behavior,  as well as the transition times between them. Through the functional dependency of the reactive mixing scale on time, we have developed a compression-diffusion equation which is able to predict the spatial localization of reactivity as a function of time, Damk{\"o}hler and P{\'e}clet numbers. We have also proposed an efficient and fast numerical scheme based on a Chebyshev spectral collocation method in order to tackle the nonlinear reaction-advection-diffusion problem numerically, which has allowed us to verify successfully all our theoretical predictions. 

While previous works had mostly focused on the cases of $Pe=0$ with variable $Da$ or $Da=\infty$ with variable $Pe$, the theoretical framework presented here spans the full space of $Pe$ and $Da$ (see figure \ref{Fig:synthesis}).
In the case of a stretching that is weak in comparison to the reaction, i.e. for $Pe < Da$, the effects of kinetics limitations and mixing enhancement by shear are essentially decoupled. Kinetics limitation is dominant in the early time regime for $t<t_{Da_1} = 32/Da$, while stretching enhanced mixing is dominant in the late time regime $t>t_{Pe_1} = 1/Pe$. In the intermediate regime reactions are limited by diffusion, leading to the classical diffusive scaling. In the case of a stretching that is strong in comparison to the reaction kinetics , i.e. for $Pe > Da$, we have shown the existence of an intermediate regime at times $t_{Pe_2}<t<t_{Da_2}$, with $t_{Pe_2} = \frac{1}{Pe}5\sqrt{3}/2$ and $t_{Da_1} = \frac{1}{Da}80\sqrt{3}$, where stretching enhanced mixing and kinetics limitations are strongly coupled. This leads to an accelerated effective kinetics with $m_c \approx \frac{Pe Da}{10\sqrt{3}}t^{5/2}$.

The presence of a background velocity gradient affects the evolution of the reactive mixing scale, which defines the region around the interface between the two reactants where reaction mostly occurs. In the absence of an imposed shear, the reactive mixing scale initially grows as $s_r \sim t^{1/2}$, due to control by molecular diffusion. Beyond the time $t_{Da_1}$ this growth slows down as $s_r \sim t^{1/6}$ due to consumption of reactants by the chemical reaction. The presence of a velocity gradient affects this evolution by introduction a regime of compression in the time evolution of the reactive mixing scale. We have attempted heuristically to formalize the nature of this evolution of the reactive mixing scale by introducing a compression-diffusion equation for each of the two different cases of weak and strong stretching. We have shown that the reactive layer is compressed between the characteristic stretching time $t\sim Pe^{-1}$  and the mixing time $t_m \sim Pe^{-2/3}$, at which diffusion balances compression. Beyond this time the reactive mixing scale evolves as $s_r \sim t^{1/6}$ in the case of weak stretching and $s_r \sim t^{1/2}$ in the case of strong stretching. The compression-diffusion equations are able to predict the nature of the evolution of the reactive mixing scale in their respective domains of validity with reasonable accuracy. 

The fundamental results presented in this work are a first step towards understanding how complex stretching processes found in particular in porous media interact with the reactivity of transported solutes to determine the dynamics of upscaled effective kinetics and the degree of spatial localization of reactive hotspots. Prospects include extending the numerical framework presented here to more complex stretching configurations, involving spatial and temporal fluctuations in stretching rates, as well as more complicated reaction kinetics involving multistep or higher order reactions.

\section{Acknowledgments}
AB would gratefully like to acknowledge post-doctoral funding through the \textit{Agence Nationale de la Recherche} (ANR-14-CE04-0003-01 subsurface mixing and reaction). 
MD acknowledges the support of the European Research Council (ERC) through the project MHetScale (617511).

\appendix
	\section{Validation of 1D methodology}
	\label{appendix:validation}
	
	\begin{figure}[h]
	\begin{centering}
\includegraphics[scale=0.6]{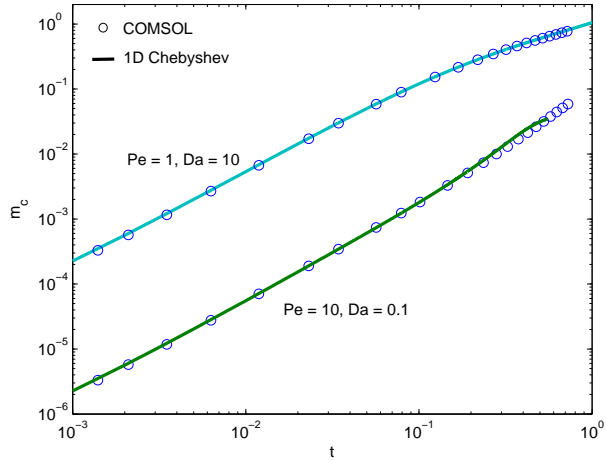}
\caption{Temporal evolution of the mass of product formed for two different cases: $Da = 0.1, Pe = 10$ and $Da = 10, Pe = 1$. The symbols denote the results obtained from the COMSOL 2D simulations, while the solid lines are those obtained from the 1D Chebyshev collocation method (please see \ref{sec:discrete})}
\label{Fig:validation}
\end{centering}
\end{figure}
	We present here a validation of the 1D Chebyshev spectral collocation method which has been employed in the present work. We compare the temporal evolution of the mass of the product formed against 2D simulations performed in the finite element framework of COMSOL Multiphysics. 	
	For the 2D simulation, we consider the case of a stretching front for two different cases, $Da = 0.1, Pe = 10$ and $Da = 10, Pe = 1$. The grid spacing employed in COMSOL is $0.1$ while the domain size is taken to be $20 \times 20$. Each front is chosen to be of size $10$, which, compared to the nondimensional length $1$ is large enough to prevent the periodic boundary condition from interacting with the diffusion front in the time of interest of our observation (we refer the reader to section \ref{sec:system_descrip} for a discussion regarding this). 
	The minimum grid size decides the initial width of the reaction front and, consequently, the system evolution. For the 1D spectral simulation presented in figure \ref{Fig:validation}, we have employed a domain length of $30$ with $N = 384$ intervals for the 1D Chebyshev spectral collocation method. The time step in the warped time system is chosen to be $\Delta\theta = 10^{-3}$.

	If we focus on the periodic boundary condition employed, we may assume that the effects of diffusion in the Lagrangian frame are primarily focused on the region near the interface. Only when the reactive mixing scale becomes large enough to overlap with the reactive mixing scales of the products formed at the periodic ends, does the assumption that $z\rightarrow\pm\infty$, which signifies the semi-infinite domain as per the 1D analysis in the present work, begins to lose its validity. Our study focuses on the behavior of the system at times before the merging of the two fronts happens.

It appears clearly in figure \ref{Fig:validation} that the temporal evolution of the mass of the product, $m_c$, obtained through the two aforementioned methods (the solution obtained through the 2D simulations in COMSOL Multiphysics is denoted by the symbols while that obtained from the 1D methodology is denoted by the solid lines) are in excellent agreement with each other for both the cases of $Da = 10, Pe = 1$ and $Da = 0.1, Pe = 10$.

\begin{figure*}[!th]
\begin{centering}
\includegraphics[scale=0.6,trim={0cm 3cm 1cm 2cm},clip]{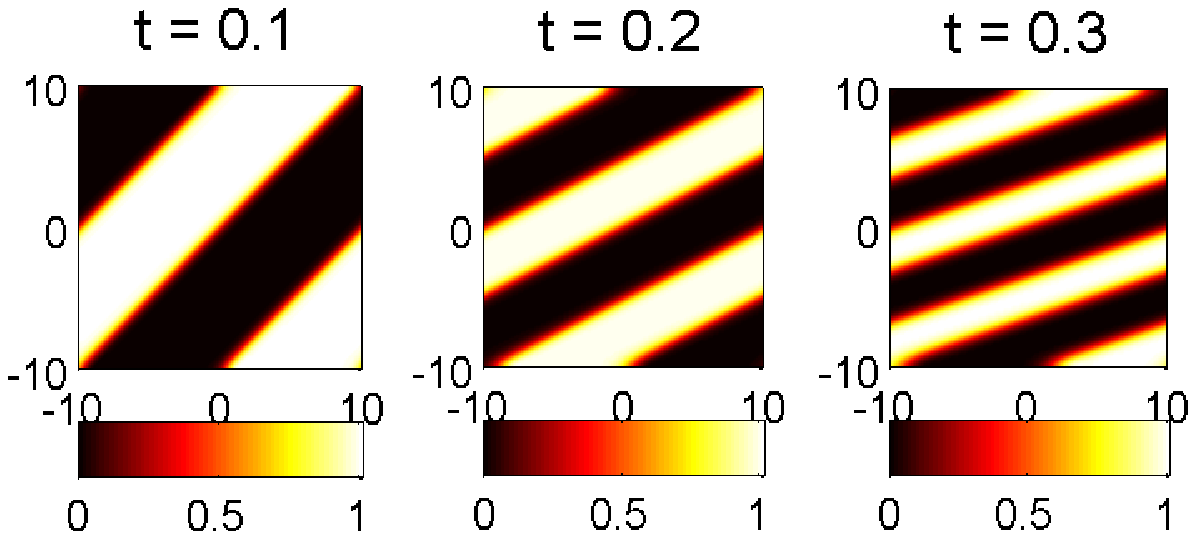}
\caption{Surface plot of the reactant concentrations at different times during the 2D COMSOL simulation. We have chosen $Pe = 10$ and $Da = 10$ in this case.}
\label{Fig:surf1}
\includegraphics[scale=0.6,trim={0cm 3cm 1cm 1cm},clip]{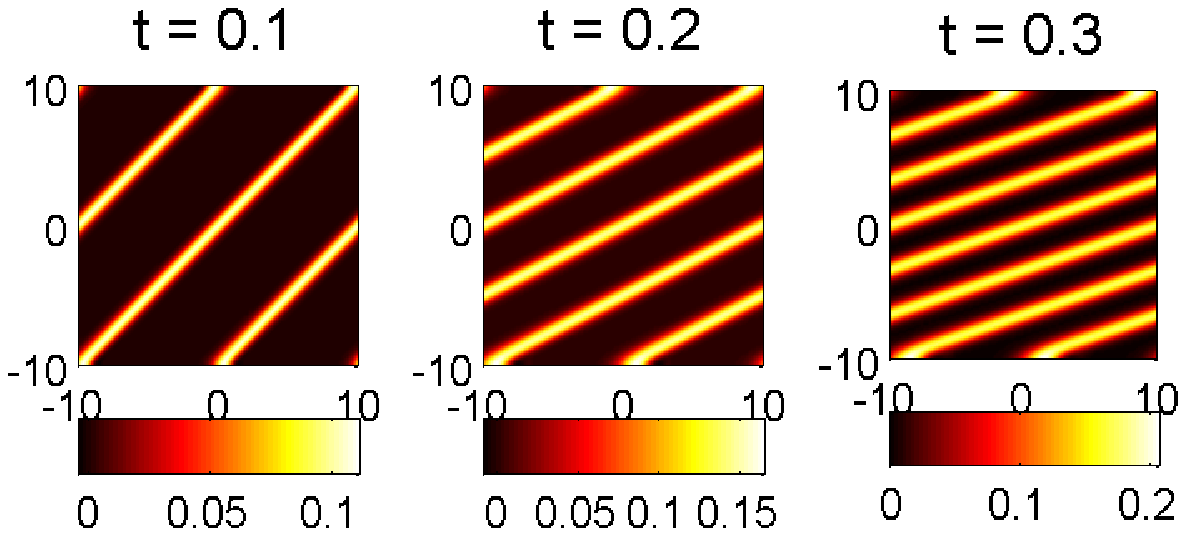}
\caption{Surface plot of the product concentration at different times during the 2D COMSOL simulation. We have chosen $Pe = 10$ and $Da = 10$ in this case.}
\label{Fig:surf2}
\end{centering}
\end{figure*}

	\begin{figure}[!th]
	\begin{centering}
\includegraphics[scale=0.6]{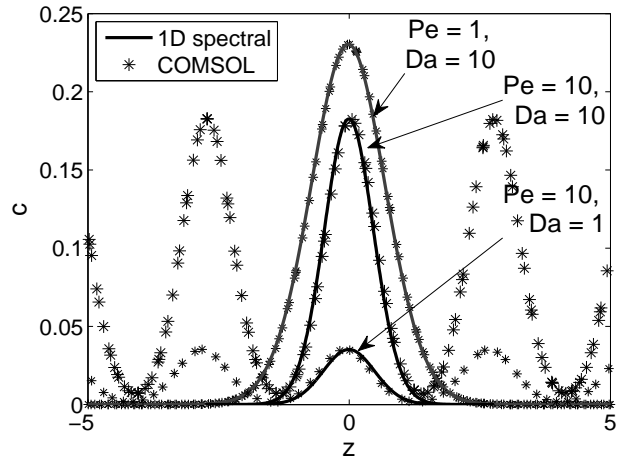}
\caption{Concentration profiles at $t = 0.35$ obtained in the direction perpendicular to the stretching interface from COMSOL 2D simulations and the 1D Chebyshev spectral collocation method. We depict the concentration profiles for 3 different cases $Pe = 1, Da = 10$, $Pe = 10, Da = 10$ and $Pe = 10, Da = 1$. The presence of multiple peaks in the figure from the 2D simulation is due to the fact that the periodicity employed for the 2D simulation causes the neighbouring fronts to enter into the region of interest (this can be confirmed from the surface plots depicted in figure \ref{Fig:surf2}.}
\label{Fig:conc}
\end{centering}
\end{figure}

In figure \ref{Fig:surf1} and \ref{Fig:surf2} we depict how the reactants and the products are stretched along the direction of the shear, leading to a corresponding compression in the perpendicular direction for times $t = 0.1, 0.2, 0.3$ for the case where $Pe = 10$ and $Da = 10$. As time progresses, the fronts from the other periodic cells start entering our cell of interest. Here at very large times, the fronts may come very close together so that the reactive mixing scales may overlap \cite{LeBorgne2013}. The product is formed at the interface between the two fronts and thus a larger number of bands can be observed for the surface plot of the product (figure \ref{Fig:surf2}).

	In figure \ref{Fig:conc}, we depict the concentration distributions obtained from the COMSOL simulations in the direction normal to the interface, and compare them with those obtained from the 1D Chebyshev spectral collocation simulation. 
It is observed that for the case of small P{\'e}clet the number of peaks at $t = 0.35$ in the domain depicted in figure \ref{Fig:conc} is only one while at a higher Péclet, we observe multiple peaks. It may be noted that so long as the concentration between the two peaks reaches zero, which implies independence of the two peaks, the results from the 1D simulation are valid. 
	It may also be confirmed from figure \ref{Fig:conc} that the concentration profiles predicted from the 1D Chebyshev spectral collocation method and full 2D simulation are in excellent agreement with each other. 
In the case of the 1D simulations, the boundary condition that is implemented is in effect that of an infinite span of the domain in either directions. In the case of 2D simulations, we have taken into account the finite length of the reaction fronts. Given that the predictions of the methodology is applicable to a time before which the periodic fronts coalesce, the scenario of the two simulations are essentially the same. 
	Hence the methodology elucidated in section \ref{sec:discrete} may be extended for complicated forms of the reactions with relative ease and with a high degree of accuracy.

\section*{References}
\bibliographystyle{elsarticle-num}
\bibliography{reaction_diffusion_v1.3}
\end{document}